\newif\ifconfver
\confverfalse      %declaring conference version false
\confvertrue        %declaring conference version true

\newif\ifplainver  %declare a plain version
\plainvertrue
\newif\ifhide  %hide something, like a  proof
\hidetrue
\ifplainver
    \confverfalse   %automatically disable conf. version argument if it's plain
\fi

\ifconfver
     \documentclass[10pt,twocolumn,twoside]{IEEEtran}
\else
    \ifplainver
        \documentclass[11pt]{article}
        \usepackage{fullpage}
    \else
        \documentclass[12pt,draftcls,onecolumn]{IEEEtran}
    \fi
\fi

\usepackage{etoolbox}%
\usepackage{xpatch}
\usepackage{blindtext}
\usepackage{tocloft}%
\newlength{\articlesectionshift}%
\let\LaTeXStandardSection\section
\let\LaTeXStandardTheSection\thesection

\let\LaTeXStandardTheSubSubSection\thesubsubsection
\let\LaTeXStandardTheParagraph\theparagraph

\makeatletter
\newcounter{titlecounter}

\xpretocmd{\maketitle}{\ifnumgreater{\value{titlecounter}}{1}}{\clearpage}{}{} % Well, this is lazy at the end ;-)
\xpatchcmd{\maketitle}{\let\maketitle\relax\let\@maketitle\relax}{\refstepcounter{titlecounter}\begingroup
  \addtocontents{toc}{\begingroup\addtolength{\cftsecindent}{-\articlesectionshift}}%
  \addcontentsline{toc}{section}{\protect{\numberline{\thetitlecounter}{\@title~ \@author}}}%
  \addtocontents{toc}{\endgroup}
}{%
  \typeout{Patching was successful}
}{%
  \typeout{patching failed}
}%

\def\@IEEEdestroythesectionargument#1{\LaTeXStandardSection{#1}}%

\xapptocmd{\maketitle}{%
\renewcommand{\thesection}{\LaTeXStandardTheSection}%
\renewcommand{\thesubsection}{\LaTeXStandardTheSubSection}%
\renewcommand{\thesubsubsection}{\LaTeXStandardTheSubSubSection}%
\renewcommand{\theparagraph}{\LaTeXStandardTheParagraph}%
}{}{}%

\@addtoreset{section}{titlecounter}

\usepackage{calc,amsfonts,amssymb,amsmath,bm,url,color,theorem,graphicx,cite}
\usepackage{psfrag,float}
\usepackage{algorithm}
\usepackage{algorithmic}
\usepackage{soul}
\usepackage{enumerate}
\usepackage{bbm}
\usepackage{shortcuts_OPT}
\usepackage{multirow}
\usepackage{subcaption}
\usepackage{subfig}
\usepackage{float}

\usepackage{pifont}%

\ifconfver \else \usepackage{titling} \fi

\usepackage{eqparbox}

\definecolor{orange}{RGB}{255,107,0}

\def\revcolor{\color{black}}  %red

\def\rev2color{\color{black}} % orange additional markup for changes of typos from July 25, 2024
\def\revR1color{\color{black}} %  blue markup for the changes in response to the reviewers' and AE's comments

\def\revMovecolor{\color{black}} % green markup for the contents moved to the main text.
\def\finalcolor{\color{black}} % final revision

\newtheorem{Lemma}{Lemma}

\newtheorem{Theorem}{Theorem}

\newtheorem{Corollary}{Corollary}

\theorembodyfont{\rmfamily}

\newcommand\bw{\ensuremath{{\bm w}}}
\newcommand\bW{\ensuremath{{\bm W}}}

\newcommand\bx{\ensuremath{{\bm x}}}
\newcommand\by{\ensuremath{{\bm y}}}

\newcommand\bH{\ensuremath{{\bm H}}}

\newcommand\be{\ensuremath{{\bm e}}}
\newcommand\bz{\ensuremath{{\bm z}}}

\newcommand\bR{\ensuremath{{\bm R}}}

\newcommand\bX{\ensuremath{{\bm X}}}
\newcommand\bZ{\ensuremath{{\bm Z}}}

\newcommand\ba{\ensuremath{{\bm a}}}

\newcommand\bA{\ensuremath{{\bm A}}}

\newcommand\bg{\ensuremath{{\bm g}}}
\newcommand\bB{\ensuremath{{\bm B}}}

\newcommand\bbeta{\ensuremath{{\bm \beta}}}

\newcommand\bd{\ensuremath{{\bm d}}}
\newcommand\bD{\ensuremath{{\bm D}}}

\newcommand\bu{\ensuremath{{\bm u}}}
\newcommand\bv{\ensuremath{{\bm v}}}

\newcommand\bsig{\ensuremath{{\bm \sigma}}}

\newcommand\bka{\ensuremath{{\bm \kappa}}}
\newcommand\ka{\ensuremath{{\kappa}}}

\newcommand\bY{\ensuremath{{\bm Y}}}

\newcommand{\Rbb}{\mathbb{R}}
\newcommand{\Cbb}{\mathbb{C}}

\newcommand{\setB}{\mathcal{B}}
\newcommand{\setD}{\mathcal{D}}

\newcommand{\setX}{\mathcal{X}}

\newcommand{\setU}{\mathcal{U}}
\newcommand{\setV}{\mathcal{V}}
\newcommand{\setW}{\mathcal{W}}
\newcommand{\setS}{\mathcal{S}}

\newcommand{\jj}{\mathfrak{j}}

\newcommand{\Diag}{\mathrm{Diag}}

\newcommand{\bzero}{{\bm 0}}
\newcommand{\bone}{{\bm 1}}
\newcommand{\bI}{{\bm I}}

\newcommand\bigO{\ensuremath{{\mathcal{O}}}}
\newcommand\half{\ensuremath{{\frac{1}{2}}}}

\newcommand\conv{\ensuremath{{\rm conv}}}

\newcommand\tr{\ensuremath{{\rm tr}}}

\newcommand\dist{\ensuremath{{\rm dist}}}
\newcommand\fro{\ensuremath{{\rm F}}}

\usepackage{tikz}
\usetikzlibrary{arrows.meta}
\usetikzlibrary{decorations}
\usetikzlibrary{calc}
\usetikzlibrary{shapes.geometric}
\usetikzlibrary{external}

\patchcmd{\maketitle}
 {\def\@makefnmark}
 {\def\@makefnmark{}\def\useless@macro}
 {}{}
\renewcommand\footnotemark{}
 
\begin{document}

%\bibliographystyle{IEEEtran}

%--- I do things quite strangely here to accommodate three style modes.
%--- input title and abstract here; it applies to all modes
%--- it's too complex to do authors or they are input for each mode
\newcommand{\papertitle}{
Extreme Point Pursuit---Part II: Further Error Bound Analysis and Applications
}

\newcommand{\paperabstract}{
In the first part of this study, a convex-constrained penalized formulation was studied for a class of constant modulus (CM) problems.
In particular, the error bound techniques were shown to play a vital role in providing exact penalization results.
In this second part of the study, we continue our error bound analysis for the cases of partial permutation matrices, size-constrained assignment matrices and non-negative semi-orthogonal matrices.
We develop new error bounds and penalized formulations for these three cases, and the new formulations
%open opportunities 
possess good structures 
for building computationally efficient algorithms. 
%than their counterparts in the first part of the study.
Moreover, we provide numerical results to demonstrate our framework in a variety of applications such as the densest $k$-subgraph problem, graph matching, size-constrained clustering, non-negative orthogonal matrix factorization and sparse fair principal component analysis.
}

%--------

\ifplainver

%    \date{May 30, 2014}

    \title{\papertitle}

    \author{
   	Junbin Liu$^\dagger$, Ya Liu$^\dagger$, Wing-Kin Ma$^\dagger$, Mingjie Shao$^\S$ and Anthony Man-Cho So$^\ddagger$\\
     \\
    $^\dagger$ Department of Electronic Engineering, The Chinese University of Hong Kong,\\
     Hong Kong SAR of China\\[.5em] 
    $^\ddagger$ Department of Systems Engineering and Engineering Management,\\ The Chinese University of Hong Kong,
    Hong Kong SAR of China\\[.5em] 
    $^\S$School of Information Science and Engineering, \\ Shandong University, Qingdao, China
%    \thanks{This work was supported by the General Research Fund (GRF) of Hong
%Kong Research Grant Council (RGC), under Project IDs CUHK 14205421 and CUHK 14208819.}
    %\thanks{The work of J. Liu, Y. Liu, W.-K. Ma and M. Shao was supported by the General Research Fund (GRF) of Hong Kong Research Grant Council (RGC) under Project ID CUHK 14208819. The work of A. M.-C. So was supported by the GRF of RGC under Project ID CUHK 14205421.} 
    \thanks{The work of J. Liu, Y. Liu, W.-K. Ma and M. Shao was supported by the General Research Fund (GRF) of Hong Kong Research Grant Council (RGC) under Project ID CUHK 14208819. The work of A. M.-C. So was supported by the GRF of Hong Kong RGC under Project ID CUHK 14205421.} 
    \thanks{Junbin Liu and Ya Liu have equal contributions.}
    }
    \maketitle

    \begin{abstract}
    \paperabstract
    \end{abstract}

    \bigskip


\else
    \title{\papertitle}

    \ifconfver \else {\linespread{1.1} \rm \fi

    \author{Junbin Liu, Ya Liu, Wing-Kin Ma, Mingjie Shao and Anthony Man-Cho So
    %\thanks{This research was supported by project \#MMT-8115059 of the Shun Hing
 %      Institute of Advanced Engineering, The Chinese University of Hong Kong. The conference version of this paper appeared in ICASSP 2018.}
     \thanks{The work of J. Liu, Y. Liu, W.-K. Ma and M. Shao was supported by the General Research Fund (GRF) of Hong Kong Research Grant Council (RGC) under Project ID CUHK 14208819. The work of A. M.-C. So was supported by the GRF of Hong Kong RGC under Project ID CUHK 14205421.} 
    \thanks{Junbin Liu and Ya Liu have equal contributions.}
    }

\maketitle

    \ifconfver \else
        \begin{center} \vspace*{-2\baselineskip}
        %11th Revision, \today \\[2\baselineskip]
        \end{center}
    \fi

    \begin{abstract}
        \paperabstract
    \end{abstract}

%    \begin{keywords}\vspace{-0.0cm}
%        ...
%    \end{keywords}

    \begin{IEEEkeywords}\vspace{-0.0cm}
        constant modulus optimization, non-convex optimization, error bound, densest subgraph problem, PCA, graph matching, clustering, ONMF
    \end{IEEEkeywords}

    \ifconfver \else \IEEEpeerreviewmaketitle} \fi

 \fi

\ifconfver \else
    \ifplainver \newpage \else        
\fi \fi

%--------------------------------------
%\newpage 

\section{Introduction}

% \revcolor causes some space, so I use negative spacing to neutralize it
In Part I of this paper{\revcolor\cite{liu2024extreme}}, we considered a convex-constrained minimization framework for a class of constant modulus (CM) problems.
Named extreme point pursuit (EXPP), this framework gives simple well-structured reformulations of the CM problems.
This allows us to apply basic methods, such as the projected gradient (or subgradient) method, to build algorithms for CM problems;
some underlying assumptions with the objective function are required, but they are considered reasonable in a wide variety of applications.
As a requirement, EXPP chooses the constraint set as the convex hull of the CM set.
When the projected gradient method is used, the computational efficiency will depend on whether the projection onto the convex hull of the CM set is easy to compute.
We examined a number of CM examples that have such a benign property.
However, we also encountered some CM sets, namely, the partial permutation matrix (PPM) set and the size-constrained assignment matrix (SAM) set, whose convex hull projections may be inefficient to compute. 
Another difficult set is the non-negative semi-orthogonal matrix (NSOM) set, whose convex hull is not even known.

%As the second part of this study, 
As Part II of this paper,
we continue our analysis with the PPM set, the SAM set, and the NSOM set.
Our focus is on the error bound principle for exact penalization, and we want to see if we can relax the EXPP constraint set for the above three cases.
We will consider some efficient-to-project constraint sets for the three cases, and we will show that, by suitably modifying the penalty function, we can once again achieve exact penalization.
These case-specific results are more powerful versions of 
%the general EXPP results 
their counterparts in Part I of this paper.
%in the first part of our study.
In the graph matching application, for example, we will numerically illustrate that the new method has much better computational efficiency than the state of the arts.
Also, our results for NSOMs draw connections to some recently emerged results \cite{wang2021clustering,jiang2023exact,chen2022tight}, as we shall see.

In addition to error bound analysis, we will numerically demonstrate EXPP in a variety of applications---such as %MIMO detection, 
the densest $k$-subgraph problem, graph matching, size-constrained clustering, ONMF, and PCA with sparsity and/or fairness.
We will see that EXPP is a working method across different applications, offering reasonable performance and computational efficiency.
%Discussion of prior related works and notations can be found in Part I of this study.
%The notations also follow those in Part I.
The organization is as follows. 
Section \ref{sect:p2_recap} recapitulates some key concepts in Part I of this study.
Section \ref{sect:p2_eb} provides further error bound analysis and devises new EXPP formulations.
Section \ref{sect:p2_num} provides numerical results for different applications.
Section \ref{sect:p2_con} concludes this paper.
%The notations used are same as those in Part I.
%We already discussed prior related works in Part I, and our development in the coming sections will cover more details with relevant prior results.

\section{A Recap of Part I}
\label{sect:p2_recap}

We give a summary of 
%the first part of our study, 
Part I of this study,
with a focus on the error bound principle.
Let $\setD \subseteq \Rbb^n$ be a set, which will be used to denote the domain of a problem.
Let $\setX \subseteq \setD$ be a set, which will often be chosen as a convex compact set.
We consider a class of CM problems in the form of 
\beq \label{eq:cm_part2}
\min_{\bx \in \setV} f(\bx),
\eeq 
where 
$f: \setD \rightarrow \Rbb$ is $K$-Lipschitz continuous on $\setX$;  
$\setV \subseteq \setX$ is a non-empty closed CM set with modulus $\sqrt{C}$.
Consider a general penalized formulation of \eqref{eq:cm_part2}:
\beq \label{eq:cm_pen_part2}
\min_{\bx \in \setX} F_\lambda(\bx):= f(\bx) + \lambda \, h(\bx),
\eeq 
where 
$h:\Rbb^n \rightarrow \Rbb$ is a penalty function; $\lambda > 0$ is a given scalar.
We want to find an $\setX$ and an $h$ such that the penalized formulation \eqref{eq:cm_pen_part2} is ``easy'' to build an algorithm and has exact penalization guarantees.  
By exact penalization, we mean that the solution set of \eqref{eq:cm_pen_part2} is equal to that of \eqref{eq:cm_part2}.
According to the error bound principle, 
%the penalized 
formulation \eqref{eq:cm_pen_part2} is an exact penalization formulation of  \eqref{eq:cm_part2} if $h$ is effectively an error bound function of $\setV$ relative to $\setX$, i.e.,
\begin{align*}
    \dist(\bx,\setV) \leq \nu \, h(\bx), & \quad \forall \bx \in \setX, \\
    0  = h(\bx), &  \quad \forall \bx \in \setV,
\end{align*}
for some constant $\nu > 0$.
More specifically, the exact penalization result holds when $\lambda$ is sufficiently large such that $\lambda > K\nu$.

We studied a convex-constrained minimization formulation of problem \eqref{eq:cm_part2}, called EXPP, which is an instance of the formulation \eqref{eq:cm_pen_part2} with
%In EXPP we choose 
\beq \label{eq:EXPP_choice}
\setX  = \conv(\setV), \quad h(\bx) = C - \| \bx \|_2^2.
\eeq 
%Such choice was motivated by the concave minimization principle for exact penalization, which was studied in the first part of this study.
% the corresponding penalized formulation in \eqref{eq:cm_pen_part2} is called EXPP.
We showed that the choice in \eqref{eq:EXPP_choice} leads to an error bound, and consequently exact penalization, for many practical CM sets of interest.
A merit with EXPP is that its penalty function $h$ in \eqref{eq:EXPP_choice} is simple.
If the constraint set $\setX$ in \eqref{eq:EXPP_choice} is friendly to handle, then we may build algorithms for EXPP without much difficulty.  
There are various possibilities for one to build algorithms for EXPP. Here we focus our attention on the projected gradient method or related methods.
Assuming that $f$ is differentiable, the projected gradient method for problem \eqref{eq:cm_pen_part2} is given by
\beq \label{eq:pgd_part2}
\bx^{l+1} = \Pi_{\setX}( \bx^l - \eta_l \nabla F_\lambda(\bx^l) ),
\quad l=0,1,\ldots,
\eeq 
where $\eta_l > 0$ is the step size.
The computational efficiency of the projected gradient method depends on whether the projection $\Pi_{\setX}$ is efficient to compute.
We examined a collection of CM sets that have efficient-to-compute projections.
However, we also encounter CM sets whose projections can be expensive, or has no known way, to compute.
\begin{enumerate}[1.]			
	\item {\em Partial permutation matrix (PPM) set:}
	\[
	\setU^{n,r} = \{ \bX \in \{ 0, 1 \}^{n \times r} \mid  \bX^\top \bone = \bone, \bX \bone \leq \bone \},
	\]
	where $n \geq r$. 
    The convex hull of $\setU^{n,r}$ is 
    \beq
	\conv(\setU^{n,r}) = \{ \bX \in [ 0, 1 ]^{n \times r} \mid  \bX^\top \bone = \bone, \bX \bone \leq \bone \}.
    \label{eq:cvx_hull_Unr}
	\eeq 
    %The projection onto $\conv(\setU^{n,r})$ has no known method to use the structure of $\conv(\setU^{n,r})$ to compute the projection efficiently.
    %There is no known specialized algorithm for computing the projection onto $\conv(\setU^{n,r})$.
    {\revR1color There is no known easy way to compute the projection onto $\conv(\setU^{n,r})$.}
    Solving the projection using 
    %a convex 
    {\revR1color a numerical}
    solver can in practice be expensive for large $n$ and $r$.
 	
	\item {\em Size-constrained assignment matrix (SAM) set:}
	\beq \label{eq:Unr_ka_part2}
	\setU^{n,r}_{\bka} = \{ \bX \in \{ 0, 1 \}^{n \times r} \mid  \bX^\top \bone = \bka, \bX \bone \leq \bone \},
	\eeq 
	where $n \geq r$, $\bka \in \{1,\ldots,n\}^r$, $\sum_{j=1}^r \ka_j \leq n$.
    The convex hull of $\setU^{n,r}_{\bka}$ is
    \begin{equation}
    \label{eq:cvx_hull_Unrk}
	\conv(\setU^{n,r}_{\bm \kappa})
	  = \{ \bX \in [ 0,1 ]^{n \times r} \mid \bX^\top \bone = \bm \kappa, \bX \bone \leq \bone \}.
	%:= \breve{\setQ}_{\bm \kappa}^{n,r}
	\end{equation}
    The projection onto $\conv(\setU^{n,r}_{\bm \kappa})$ has the same issue as that in the PPM case.
 %    For any $\bX \in \conv(\setU^{n,r}_{\bm \kappa})$ we have the error bounds
 %    \begin{subequations} \label{eq:error_Unrk}
	% 	\begin{align}
	% 		\dist(\bX,\setU^{n,r}_\bka) & \leq 
 %            %3 \sqrt{\bone^\top \bka} 
	% 		% \textstyle 
 %            \nu 
	% 		\left[ \sum_{j=1}^r (\ka_j - s_{\ka_j}(\bx_j) ) \right]  \\
	% 		& \leq 
 %            \nu
 %            %3 \sqrt{\bone^\top \bka} 
 %            ( \bone^\top \bka - \| \bX \|_{\rm F}^2 ),
	% 	\end{align}
	% \end{subequations}
 %    where $\nu = 3 \sqrt{\bone^\top \bka} $.
 %    For the last case $\setU^{n,r}$, we have the same error bounds with $\nu = 3 \sqrt{r}$ and $\bka = \bone$.
	
	\item {\em Non-negative  semi-orthogonal matrix (NSOM) set:} 
	$$\setS^{n,r}_+ = \setS^{n,r} \cap \Rbb_+^{n \times r},$$ where $n \geq r$
    and $\setS^{n,r} = \{ \bX \in \Rbb^{n \times r} \mid \bX^\top \bX = \bI \}$ is the semi-orthogonal matrix 
    % (SOM) 
    set.
    There are no known expressions with the convex hull of $\setS^{n,r}_+$ and the associated projection.
    %We will study its error bounds.
    %We considered the alternative $\setX = \setB^{n,r}_+$, the non-negative unit spectral norm ball, and showed an inexact penalization result for EXPP.
	
\end{enumerate}
In this second part of the paper, we continue to study these CM sets.

\section{Further Error Bound Analysis}
\label{sect:p2_eb}

In this section we perform error bound analysis for the 
above three sets.
%partial permutation matrix set $\setU^{n,r}$, the size-constrained assignment matrix set $\setU^{n,r}_{\bka}$, and the non-negative semi-orthogonal matrix set $\setS_+^{n,r}$.
The first and second subsections consider the PPM case.
We will provide an application example to motivate the study, and then we will derive a new error bound and EXPP to overcome the computational issue described in the last section.
Following the same genre, 
the SAM case is tackled in the third and fourth subsections,
and the NSOM case studied in the fifth, sixth, and seventh subsections.

\subsection{Example: Graph Matching}
\label{sect:exa_gm}

Consider the following problem for a given $\bA, \bB \in \Rbb^{n \times n}$:
\beq
\min_{\bX \in \setU^{n,n}} \|  \bA - \bX^\top \bB \bX  \|_{\fro}^2.
\nonumber
\eeq 
This is called the graph matching (GM) problem in the context of computer vision.
The goal is to match the nodes of two equal-size graphs by using the graph edge information.
The GM problem does so by finding a set of one-to-one node associations, represented by $\bX$, such that the associated Euclidean error between the two graphs' adjacency matrices, represented by $\bA$ and $\bB$, is minimized.
Since a feasible $\bX$ is orthogonal, the GM problem can be rewritten as
\beq \label{eq:prob_gm}
\min_{\bX \in \setU^{n,n}} f(\bX) := \| \bX \bA - \bB \bX \|_{\fro}^2.
\eeq
Following the review in the last section, the EXPP formulation of \eqref{eq:prob_gm} is
\beq \label{eq:expp_prob_gm}
\min_{\bX \in \conv(\setU^{n,n})} F_\lambda(\bX) = f(\bX) - \lambda \| \bX \|_{\fro}^2.
\eeq
%where $\conv(\setU^{n,n})$ is given by \eqref{eq:cvx_hull_Unr}.
%While this is a convex-constrained minimization problem,
It can be 
expensive 
%{\revR1color computationally burdensome}
to apply the projected gradient method \eqref{eq:pgd_part2} to 
the EXPP--GM problem 
\eqref{eq:expp_prob_gm}.
{\revR1color The projection operation $\Pi_{\conv(\setU^{n,n})}$ in the projected gradient method requires us to solve an optimization problem,
namely,
minimization of $\| \bZ - \bX \|_{\fro}^2$ over $\bX \in \conv(\setU^{n,n})$, where $\bZ$ is given.
As mentioned, there is no easy-to-compute solution for this problem.
The problem is nevertheless convex, and there is a specialized numerical solver for this problem based on gradient ascent of the dual problem \cite{jiang2016l_p}.
Still, one would anticipate that solving an optimization problem at each step of the projected gradient method \eqref{eq:pgd_part2} would be computationally burdensome particularly when the problem size $n$ is large.
}

%As mentioned previously,
%it can be costly to compute the projection $\Pi_{\conv(\setU^{n,n})}$ when $n$ is large.
% there is no known way to exploit the structure of $\conv(\setU^{n,n})$ to efficiently compute the projection $\Pi_{\conv(\setU^{n,n})}$,
% and solving the projection using a convex solver is, in practice, expense for large $n$.
% This stands as a hurdle computationally.
%to apply the projected gradient method \eqref{eq:pgd_part2} to the EXPP-GM problem \eqref{eq:expp_prob_gm}.

At this point it is worthwhile to review an important prior study in GM \cite{zaslavskiy2008path}.
The authors in that study developed a similar formulation as \eqref{eq:prob_gm};
they essentially put forth the same fundamental idea as EXPP under the concave minimization principle, though
not under the error bound principle.
They built an algorithm that exploits the problem structure for efficient computations. 
Specifically they considered the Frank-Wolfe method, which for \eqref{eq:expp_prob_gm} is given by
\beq
\bX^{l+1} = \bX^l + \alpha_l ( {\sf LO}_{\conv(\setU^{n,n})}( \nabla F_\lambda(\bX^l) ) - \bX^l ),
\nonumber 
\eeq 
where $\alpha_l \in [0,1]$ is a step size;
${\sf LO}_\setX(\bg) \in \arg \min_{\bx \in \setX} \bg^\top \bx$ is the linear optimization 
{\rev2color (LO)}
oracle for $\setX$.
To efficiently compute each Frank-Wolfe iteration, we need an efficient way to compute ${\sf LO}_{\conv(\setU^{n,n})}$.
The authors of \cite{zaslavskiy2008path} did so by using the Hungarian algorithm \cite{mcginnis1983implementation},
which is a specialized algorithm for solving linear optimization over the set of doubly stochastic matrices and can efficiently compute ${\sf LO}_{\conv(\setU^{n,n})}$ with a complexity of  $\bigO(n^3)$.

We have a different proposition. 
To put into context, recall that $\Delta^n = \{ \bx \in \Rbb^n_+ \mid \bx^\top \bone = 1 \}$ denotes the unit simplex.
Define 
\ifconfver
\begin{align*}
    \tilde{\Delta}^{n \times r} & = \{ \bX \in [0,1]^{n \times r} \mid \bX^\top \bone = \bone \} 
    \\
    & 
    = \{ \bX \in \Rbb^{n \times r} \mid \bx_j \in \Delta^n, ~ \forall j \}.
    \nonumber 
\end{align*}
\else
\begin{align*}
    \tilde{\Delta}^{n \times r} & = \{ \bX \in [0,1]^{n \times r} \mid \bX^\top \bone = \bone \}
    = \{ \bX \in \Rbb^{n \times r} \mid \bx_j \in \Delta^n ~ \forall j \},
    \nonumber 
\end{align*}
\fi
This set is a relaxation of $\conv(\setU^{n,r})$ 
in \eqref{eq:cvx_hull_Unr} by taking out the row constraint $\bX \bone \leq \bone$.
Our idea is to derive an error bound function $h$ of $\setU^{n,r}$ relative to $\tilde{\Delta}^{n \times r}$.
%(not $\conv(\setU^{n,r})$).
If we can do so, then we will have an exact penalization formulation \eqref{eq:cm_pen_part2} for the GM problem, and more generally, CM problems 
%\eqref{eq:cm_part2} 
for the PPM case.
%Suppose that the resulting formulation \eqref{eq:cm_pen_part2} has a differentiable objective function $F_\lambda$.
Consequently, there is a possibility for us to apply the projected gradient method \eqref{eq:pgd_part2} in a computationally efficient fashion.
%than that for the EXPP formulation (e.g., \eqref{eq:expp_prob_gm} for GM).
To be specific, we can compute the projection $\Pi_{\tilde{\Delta}^{n \times r}}(\bZ)$ of a given matrix $\bZ \in \Rbb^{n \times r}$ with a complexity of $\bigO(n r \log(n))$:
The projection $\Pi_{\tilde{\Delta}^{n \times r}}(\bZ)$ corresponds to the unit simplex projections $\Pi_{{\Delta}^n}(\bz_j)$'s for $j=1,\ldots,r$, and the unit simplex projection can be computed with a complexity of $\bigO(n \log(n))$ \cite{condat2016fast}.
In the GM application, the complexity of the projection ${\finalcolor \Pi_{\tilde{\Delta}^{ n \times n}}}$ is $\bigO(n^2 \log(n))$---which seems more attractive than its Frank-Wolfe counterpart, $\bigO(n^3)$.

{\revR1color 
To illustrate the efficiencies of the above described projection and LO operations, we ran a numerical experiment.
We tested the runtimes of 
(i) the projection onto $\conv(\setU^{n,n})$, implemented either by CVX or by the specialized dual gradient method \cite{jiang2016l_p};
(ii) the LO oracle for $\conv(\setU^{n,n})$, implemented by the Hungarian algorithm;
and 
(iii) the projection onto $\tilde{\Delta}^{n \times n}$, done by column-wise unit-simplex projection.
The results are shown in Fig.~\ref{fig:projections_runtime}.
It is seen that the projection onto $\tilde{\Delta}^{n \times n}$ is much faster than the other operations.
}

\ifconfver
\begin{figure}[h]
    \centering
    \includegraphics[scale=0.58]
    {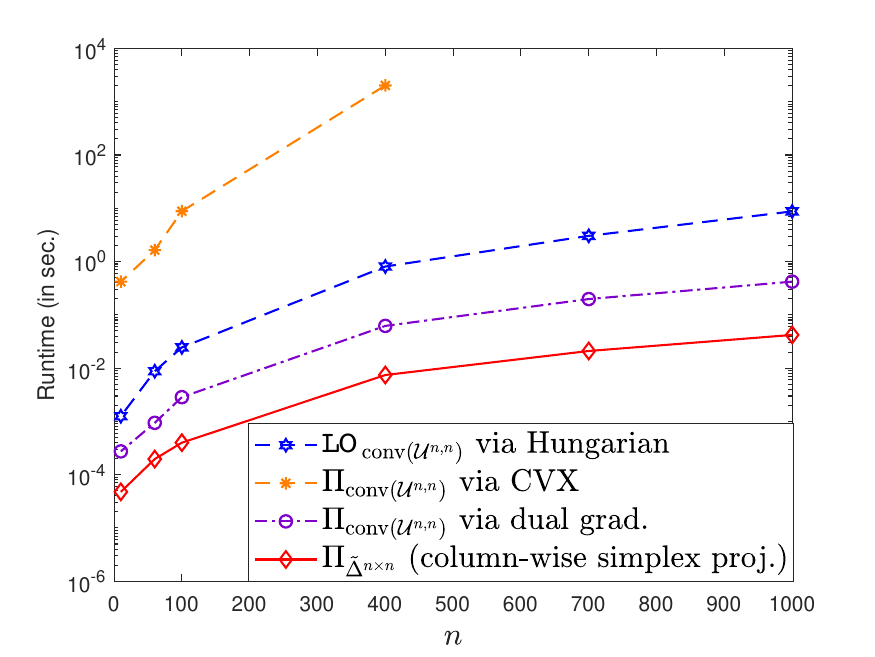}
    \caption{{\revR1colorRuntime comparison with the projections and LO oracle for the full PPM case ($n=r$).
    We used $200$ randomly generated trials to evaluate the average runtimes, except for $\Pi_{\conv(\setU^{n,n})}$ via CVX which we tested only $20$ trials due to the long runtime.}
    }
    \label{fig:projections_runtime}
\end{figure}
\else
\begin{figure}[h]
    \centering
    \includegraphics[scale=0.65]
    {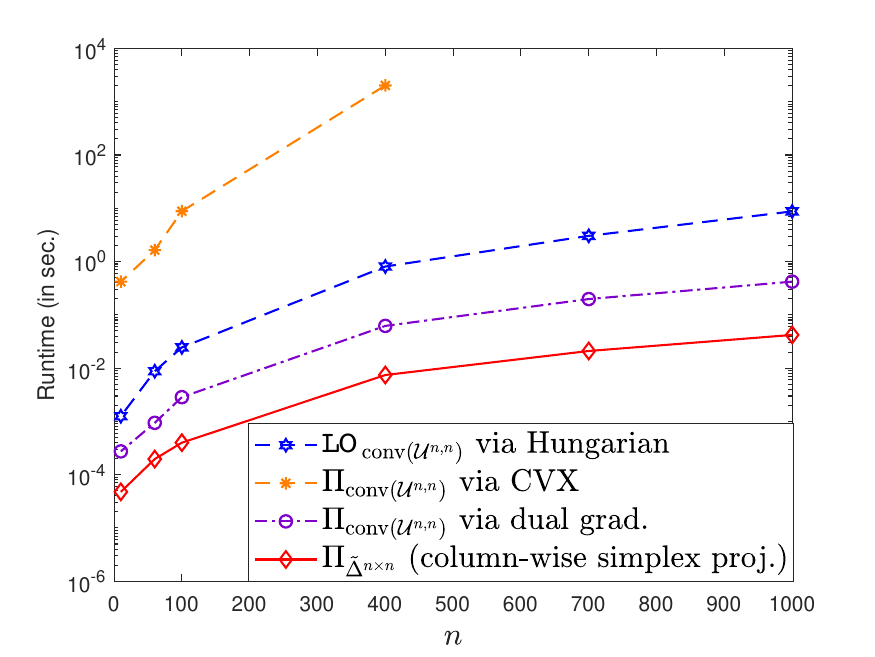}
    \caption{Runtime comparison with the projections and LO oracle for the full PPM case ($n=r$).
    We used $200$ randomly generated trials to evaluate the average runtimes, except for $\Pi_{\conv(\setU^{n,n})}$ via CVX which we tested only $20$ trials due to the long runtime.
    }
    \label{fig:projections_runtime}
\end{figure}
\fi

\subsection{A New Error Bound and a New EXPP for PPMs}
\label{sect:eb_ppm}

Motivated by the GM problem, we perform error bound analysis for the PPM set $\setU^{n,r}$ relative to the column-wise unit simplex $\tilde{\Delta}^{n \times r}$.
Our result is as follows.
\begin{Theorem} \label{thm:eb2_Unr}
	{\bf (error bound for partial permutation matrices)}
	%For any $\bX \in [0,1]^{n \times r}$ with $\bX^\top \bone = \bone$ (and without $\bX \bone \leq \bone$), 
    For any $\bX \in \tilde{\Delta}^{n \times r}$,
    we have the error bound 
    % \beq
    %     \dist(\bX,\setU^{n,r}) \leq \nu \, q(\bX),
    % \eeq 
    	\begin{subequations} \label{eq:eb2_Unr}
	\begin{align} 
		\label{eq:eb2_Unr_a}
		\dist(\bX,\setU^{n,r}) & \leq  \nu \| \bX^\top \bX - \bI \|_{\ell_1} \\
%		& = \nu  \left[ \sum_{j=1}^r (1- \| \bx_j \|_2^2) + \sum_{i=1}^n ( (\bone^\top \bar{\bx}_i)^2 - \| \bar{\bx}_i \|_2^2)      \right] \\
		\label{eq:eb2_Unr_b}
		& = \nu ( r + \| \bX \bone \|_2^2 - 2 \| \bX \|_{\rm F}^2),
	\end{align}
	\end{subequations}
	where $\nu = 11 \sqrt{r}$.
\end{Theorem}
The proof of Theorem~\ref{thm:eb2_Unr} uses the same approach as that for the error bound for $\setU^{n,r}_\bka$ 
%in the first part of this paper, Section IV.F.
in Part I, Section IV.F, of this paper.
Nevertheless, the latter is considered easier to show because of the presence of the row constraint $\bX \bone \leq \bone$.
To derive \eqref{eq:eb2_Unr} we need to go further, analyzing the singular values of $\bX$.
In view of its complexity, the proof of Theorem~\ref{thm:eb2_Unr} is relegated to Appendix \ref{app:thm:eb2_Unr}.

Theorem~\ref{thm:eb2_Unr} gives rise to a new EXPP formulation for $\setU^{n,r}$.
%Let us describe the utility of Theorem~\ref{thm:eb2_Unr}.
By applying the error bound \eqref{eq:eb2_Unr_b} to the penalized formulation \eqref{eq:cm_pen_part2}, we have an exact penalization formulation for the PPM case:
\beq \label{eq:new_expp_ppm}
\min_{\bX \in \tilde{\Delta}^{n \times r}} F_\lambda(\bX)= f(\bX) + \lambda( \| \bX \bone \|_2^2 - 2 \| \bX \|_\fro^2 ).
\eeq
%for CM problems for the partial permutation matrix case.
This new EXPP formulation resembles our previous EXPP, with addition of a friendly penalty term $\| \bX \bone \|_2^2$.
Assuming a differentiable $f$, the new EXPP problem \eqref{eq:new_expp_ppm} can be efficiently handled by the projected gradient method \eqref{eq:pgd_part2}.
%in a computational efficient fashion.
In particular, without counting the complexity of computing the gradient $\nabla F_\lambda$, the per-iteration complexity of the projected gradient method is $\bigO(n r \log(n))$; see the discussion in the last subsection.

We should provide insight into the error bound in Theorem~\ref{thm:eb2_Unr}.
%Recall that $\setS^{n,r} = \{ \bX \in \Rbb^{n \times r} \mid \bX^\top \bX = \bI \}$ denotes the semi-orthogonal matrix set.
The PPM set can be characterized as
\[
\setU^{n,r} = \tilde{\Delta}^{n \times r} \cap \setS^{n,r}.
\]
The error bound \eqref{eq:eb2_Unr_a} reveals that, by using $\tilde{\Delta}^{n \times r}$ as the constraint set and $\| \bX^\top \bX - \bI \|_{\ell_1}$ as the penalty, we can achieve exact penalization results.
Particularly, $\| \bX^\top \bX - \bI \|_{\ell_1}$ appears as a penalty for promoting $\bX$ to be semi-orthogonal.
%for promoting semi-orthogonality.
Moreover, the error bound \eqref{eq:eb2_Unr_b} is an equivalent form of $\| \bX^\top \bX - \bI \|_{\ell_1}$.
Consider the following result.
\begin{Lemma} \label{lem:q_identity}
    % For any $\bX \in \tilde{\setB}_+^{n \times r}$, the function $q$ in \eqref{eq:q_def2} equals
    % \[
    % q(\bX) = \| \bX^\top \bX - \bI \|_{\ell_1}.
    % \]
    Let $\bd \in \Rbb_{++}^r$ be given.
    Let $\bD = \Diag(\bd)$.
    For any $\bX \in \Rbb_+^{n \times r}$ with $\| \bx_j \|_2^2 \leq d_j$ for all $j$, it holds that 
    \[
    \| \bX^\top \bX - \bD \|_{\ell_1} 
    = \bone^\top \bd + \| \bX \bone \|_2^2 - 2 \| \bX \|_\fro^2.
    \]
\end{Lemma}
%The proof of Lemma~\ref{lem:q_identity} is simple and is shown as follows.

\medskip
{\em Proof of Lemma~\ref{lem:q_identity}:} \
For any $\bX \in \Rbb_+^{n \times r}$ with $\| \bx_j \|^2_2 \leq d_j$ for all $j$,
\begin{align*}
	\| \bX^\top \bX - \bD \|_{\ell_1} & = 
	\sum_{j=1}^r ( d_j - \| \bx_j \|_2^2 ) + \sum_{j=1}^r \sum_{\substack{k=1 \\ j \neq k}}^r \bx_j^\top \bx_k \\
	& =  
	\sum_{j=1}^r ( d_j - 2 \| \bx_j \|_2^2 ) + \left( \sum_{j=1}^r \bx_j \right)^\top \left( \sum_{j=1}^r \bx_j \right) \\
	& = \bone^\top \bd - 2 \| \bX \|_{\rm F}^2 + \| \bX \bone \|_2^2.
\end{align*}
%For any $\bX \in \tilde{\setB}_+^{n \times r}$, we have 
% \begin{align*}
% 	\| \bX^\top \bX - \bI \|_{\ell_1} & = 
% 	\sum_{j=1}^r ( 1 - \| \bx_j \|_2^2 ) + \sum_{j=1}^r \sum_{\substack{k=1 \\ j \neq k}}^r \bx_j^\top \bx_k \\
% 	& =  
% 	\sum_{j=1}^r ( 1 - 2 \| \bx_j \|_2^2 ) + \left( \sum_{j=1}^r \bx_j \right)^\top \left( \sum_{j=1}^r \bx_j \right) \\
% 	& = r - 2 \| \bX \|_{\rm F}^2 + \| \bX \bone \|_2^2.
% \end{align*}
The proof is complete.
\hfill $\blacksquare$
\medskip

Applying Lemma~\ref{lem:q_identity} to \eqref{eq:eb2_Unr_a} gives the error bound \eqref{eq:eb2_Unr_b}; note that any $\bx \in \Delta^n$ has $\| \bx \|_2 \leq \| \bx \|_1 = \bone^\top \bx  = 1$.
The error bound \eqref{eq:eb2_Unr_b} also reveals interesting insight.
For any $\bX \in \Rbb_+^{n \times r}$ with $\| \bx_j \|_2 \leq 1$ for all $j$, we can rewrite \eqref{eq:eb2_Unr_b} as 
\beq \label{eq:penalty_ppm_interpret}
r + \| \bX \bone \|_2^2 - 2 \| \bX \|_{\fro}^2 =
\sum_{j=1}^{\revR1color r} c_2(\bx_j) + \sum_{i=1}^n \rho_2(\bar{\bx}_i),
\eeq
where 
% \begin{align}
%     c_2(\bx) & = 1 - \| \bx \|_2^2, \\
%     \rho_2(\bx) & = | \bone^\top \bx |^2 - \| \bx \|_2^2
% \end{align}
\[
c_2(\bx) = 1 - \| \bx \|_2^2,
\quad 
\rho_2(\bx) = \| \bx \|_1^2 - \| \bx \|_2^2
\]
%The functions $c_2$ and $\rho_2$ 
appear as penalties for the columns and rows, respectively.
According to 
%the first part of this study, 
Part I of this paper,
$c_2$ is effectively an error bound function for the unit vector set $\setU^n = \{ \be_1,\ldots,\be_n \}$ relative to the unit simplex $\Delta^n$.
This means that $c_2$ promotes every column $\bx_j$ to lie in $\setU^n$.
As for $\rho_2$, we note the basic norm result that $\| \bx \|_1 \geq \| \bx \|_2$ for any $\bx$, and that $\| \bx \|_1 = \| \bx \|_2$ if and only if $\bx$ is a scaled unit vector, i.e., $\bx = \alpha \be_i$ for some $\alpha$ and $i$.
This means that $\rho_2$ promotes every row $\bar{\bx}_i$ to be a scaled unit vector.
Putting the column and row penalties together, we have the interpretation that 
%$\bX$ is promoted to be a partial permutation matrix.
\eqref{eq:penalty_ppm_interpret} promotes $\bX$ to be a PPM.
%Intuitively, this makes sense.

% Define 
% \[
% \tilde{\setB}_+^{n \times r} = \{ \bX \in \Rbb_+^{n \times r} \mid \| \bx_j \|_2 \leq 1 ~ \forall j \}.
% \]

% Hence, for $\bX \in \tilde{\setB}_+^{n \times r}$, $q(\bX)$ encourages $\bX$ to take the form
% \[
% \bX^\top = \begin{bmatrix}
%     \alpha_1 \be_{l_1}, \ldots, \alpha_n \be_{l_n}
% \end{bmatrix},
% \]
% where $\alpha_i \geq 0$ for all $i$, $\sum_{i: l_i= j} \alpha_i^2 = 1$ for all $j$.

\subsection{Example: Size-Constrained Clustering}
\label{sect:clustering_size}

We turn our interest to the SAM set $\setU_\bka^{n,r}$ in \eqref{eq:Unr_ka_part2}.
Let $\bka \in \{1,\ldots,n\}^r$, $\sum_{j=1}^r \ka_j = n$ be given.
Given a matrix $\bY \in \Rbb^{m \times n}$, we consider the following problem
\beq \label{eq:prob_scs0}
\min_{\bA \in \Rbb^{m \times r}, \bX \in \setU_\bka^{n,r}} \| \bY - \bA \bX^\top \|_\fro^2.
\eeq 
This problem appears in size-constrained clustering and was used, for instance, in paper-to-session assignment \cite{sidiropoulos2015signal}.
In size-constrained clustering, we want to cluster a given set of data points $\by_1,\ldots,\by_n$ into $r$ clusters, and the constraint is that each cluster has size $\ka_j$.
In the size-constrained clustering problem \eqref{eq:prob_scs0},
the $j$th column $\ba_j$ of 
%the decision variable 
$\bA$ describes the cluster center of cluster $j$.
From \eqref{eq:Unr_ka_part2} it is seen that 
%the decision variable 
$\bX$ takes the form
\[
\bX^\top = \begin{bmatrix}
\be_{l_1}, \ldots, \be_{l_n}
\end{bmatrix},
\]
for some $l_i \in \{ 1,\ldots,r\}$.
In particular, $\be_j$ is constrained to appear in the rows of $\bX$ for $\ka_j$ times.

A natural way to handle the size-constrained clustering problem \eqref{eq:prob_scs0} is to apply alternating minimization; see, e.g., \cite{sidiropoulos2015signal}.
In this study, we use the following reformulation of the size-constrained clustering problem
\beq \label{eq:prob_scs}
\min_{\bX \in \setU_\bka^{n,r}} f(\bX):= - \tr( \bD^{-1} \bX^\top \bR \bX ),
\eeq 
where $\bD = \Diag(\bka)$; $\bR = \bY^\top \bY$.
Problem \eqref{eq:prob_scs} is obtained by substituting the solution to $\bA$ given an $\bX \in \setU_\bka^{n,r}$, 
specifically, $\bA = \bY (\bX^\top)^\dag = \bY \bX \bD^{-1}$, to problem \eqref{eq:prob_scs0}.
Here, $\dag$ denotes the pseudo-inverse.
We encounter the same difficulty as the PPM case in Section~\ref{sect:exa_gm}:
The EXPP formulation can be applied to \eqref{eq:prob_scs}, but the projected gradient method for EXPP can be expensive to implement due to 
%the complexity of computing $\Pi_{\conv(\setU_\bka^{n,r})}$.
{\rev2color the computational cost of numerically solving $\Pi_{\conv(\setU_\bka^{n,r})}$.}
We want to replace the EXPP constraint set $\conv(\setU_\bka^{n,r})$ with
\ifconfver
\begin{align*}
    \tilde{\setU}_\bka^{n,r} & = \{ \bX \in [0,1]^{n \times r} \mid \bX^\top \bone = \bka \}
    \\
    & 
    = \{ \bX \in \Rbb^{n \times r} \mid \bx_j \in \conv(\setU_{\ka_j}^n), ~ \forall j \}. 
    \nonumber 
\end{align*}
\else
\begin{align*}
    \tilde{\setU}_\bka^{n,r} & = \{ \bX \in [0,1]^{n \times r} \mid \bX^\top \bone = \bka \}
    = \{ \bX \in \Rbb^{n \times r} \mid \bx_j \in \conv(\setU_{\ka_j}^n), ~ \forall j \}.
    \nonumber 
\end{align*}
\fi
Recall that $\setU_\ka^n = \{ \bx \in \{ 0,1 \}^n \mid \bone^\top \bx = \ka \}$ and $\conv(\setU_\ka^n) = \{ \bx \in [0,1]^n \mid \bone^\top \bx = \ka \}$, where $\ka \in \{1,\ldots,n\}$; and that $\Pi_{\conv(\setU_\ka^n)}$ can be efficiently computed by a bisection algorithm with a complexity of $\bigO(n \log(n))$ for a given solution precision \cite[Algorithm 1]{konar2021exploring}.

\subsection{A New Error Bound and a New EXPP for SAMs}

We have the following result.
\begin{Theorem} \label{thm:eb2_Unrk}
	%[An error bound for size-constrained assignment matrices]
	{\bf (error bound for size-constrained assignment matrices)}
	For any $\bX \in \tilde{\setU}_\bka^{n,r}$, we have the error bound 
	% \begin{equation} \label{eq:eb2_Unrk}
	% 		\dist(\bX,\setU^{n,r}_\bka)
	% 		 \leq \nu ( \bone^\top \bka + \| \bX \bone \|_2^2 - 2 \| \bX \|_{\rm F}^2)
	% \end{equation}	
    \begin{subequations} \label{eq:eb2_Unrk}
        \begin{align}        
        \dist(\bX,\setU^{n,r}_\bka)
			 & \leq \nu \| \bX^\top \bX - \Diag(\bka) \|_{\ell_1} \label{eq:eb2_Unrk_a} \\
            & = \nu ( \bone^\top \bka + \| \bX \bone \|_2^2 - 2 \| \bX \|_{\rm F}^2), \label{eq:eb2_Unrk_b}
        \end{align} 
    \end{subequations}
	where $\nu = 3 \sum_{i=1}^r (1+ 2\ka_i)( 1 + \sqrt{\ka_i} ) \sqrt{\bone^\top \bka}.$ 
	Note that $\nu \leq  18 \ka_{\rm max}^2 r^{3/2}$, where $\ka_{\rm max} = \max\{ \ka_1,\ldots,\ka_r \}$.
%	In particular, the constant $\nu$ is bounded by $\nu \leq 18 \ka_{\rm max}^2 r^{3/2}$ where $\ka_{\rm max} = \max\{ \ka_1,\ldots,\ka_r \}$.
\end{Theorem}
The idea behind the proof of Theorem \ref{thm:eb2_Unrk} is the same as that of the PPM case in Theorem \ref{thm:eb2_Unr}.
The actual proof of Theorem \ref{thm:eb2_Unrk} is however more tedious.
The reader can find the proof in Appendix \ref{app:thm:eb2_Unrk}.
The error bound \eqref{eq:eb2_Unrk} shares the same insight as its PPM counterpart (see  Section \ref{sect:eb_ppm}), and we shall not repeat.

Applying the error bound \eqref{eq:eb2_Unrk_b} to the penalized formulation \eqref{eq:cm_pen_part2} gives rise to the following new EXPP formulation for the SAM case:
\beq \label{eq:new_expp_sam}
\min_{\bX \in \tilde{\setU}_\bka^{n,r}} F_\lambda(\bX)= f(\bX) + \lambda( \| \bX \bone \|_2^2 - 2 \| \bX \|_\fro^2 ).
\eeq
This new EXPP is identical to the new EXPP for the partial permutation matrix case in \eqref{eq:new_expp_ppm}.
Assuming a differentiable $f$,
we can use the projected gradient method \eqref{eq:pgd_part2} to efficiently handle problem \eqref{eq:new_expp_sam}.
Specifically, the projection {\revcolor $\Pi_{\tilde{\setU}_\bka^{n,r}}$}, which contributes to the bulk of complexity with the projected gradient method, can be done with a complexity of $\bigO(n r \log(n))$;
the idea is the same as that of the PPM case (see the last paragraph of Section~\ref{sect:exa_gm}).

\subsection{Example: Orthogonal Non-Negative Matrix Factorization}
\label{sect:ONMF}

As an application example for the NSOM set $\setS_+^{n,r}$,
consider the orthogonal non-negative matrix factorization (ONMF) problem
\beq \label{eq:prob_onmf0}
\min_{\bA \in \Rbb^{m \times r}_+, \bX \in \setS_+^{n,r}} \| \bY - \bA \bX^\top \|_\fro^2,
\eeq 
where the given matrix $\bY$ is non-negative.
The ONMF problem is, in essence, a non-negatively scaled clustering problem.
It is known by researchers that any point $\bX$ in 
%the non-negative semi-orthogonal matrix set 
$\setS^{n,r}_+$ 
can be characterized as 
\beq \label{eq:nsom_char}
\bX^\top = \begin{bmatrix}
    \alpha_1 \be_{l_1}, \ldots, \alpha_n \be_{l_n}
\end{bmatrix},
\eeq 
where $\alpha_i \geq 0$ for all $i$, $l_i \in \{1,\ldots,r\}$, $\sum_{i: l_i= j} \alpha_i^2 = 1$ for all $j$;
see, e.g., 
\cite{pompili2014two,wang2021clustering,jiang2023exact,chen2022tight}.
From this characterization, we see that, fixing an assignment $l_1,\ldots,l_n$, each $\ba_j$ is a cluster center, obtained by minimizing $\sum_{i: l_i = j} \| \by_i - \alpha_i \ba_j \|_2^2$ over the non-negative $\ba_j$ and the non-negative scale-compensating scalars $\alpha_i$'s.
%In particular, the $\alpha_i$'s play the role of compensating the scaling with the $\by_i$'s.
We consider the following reformulation of the ONMF problem
\beq
 \label{eq:prob_onmf}
\min_{\bX \in \setS_+^{n,r}} f(\bX) := -\tr( \bX^\top \bR \bX),
\eeq 
where $\bR = \bY^\top \bY$.
Problem \eqref{eq:prob_onmf} is obtained by putting the solution to $\bA$ given an $\bX \in \setS^{n,r}_+$ and a $\bY \in \Rbb^{m \times n}_+$, i.e., $\bA = \bY (\bX^\top)^\dag = \bY \bX$, to \eqref{eq:prob_onmf0}.
%it can be verified that, for a given $\bY \in \Rbb^{m \times n}_+$, the solution to problem \eqref{eq:prob_onmf0} given an $\bX \in \setS^{n,r}_+$ is $\bA = \bY (\bX^\top)^\dag = \bY \bX$, and putting this solution into \eqref{eq:prob_onmf0} gives \eqref{eq:prob_onmf}.
As discussed in 
%the first part of this study, 
Part I of this paper,
there is no known expression for the convex hull of $\setS_+^{n,r}$.
%and we do not know how to apply EXPP in the non-negative semi-orthogonal matrix case.
%We hence turn to the error bound principle, seeing if we can obtain an error bound relative to a friendly constraint set. %\setX$ that is similar to, but not the same as, $\conv(\setS_+^{n,r})$.

We turn to the error bound principle, seeing if we can obtain an error bound of $\setS_+^{n,r}$ relative to a friendly constraint set $\setX$.
We should mention a recently-emerged related work \cite{jiang2023exact}.
Consider the following result.
%by Jiang {\em et al.}.
\begin{Theorem}
	%[Theorem 2 in \cite{chen2022tight}] 
	\label{thm:jiang_bnd}
	{\bf (a special case of Lemma 3.1 in \cite{jiang2023exact})}
    Define $\tilde{\setS}_+^{n \times r} = \{ \bX \in \Rbb_+^{n \times r} \mid \| \bx_j \|_2 = 1 ~ \forall j \}$
    as the column-wise non-negative unit sphere.
	For any $\bX \in \tilde{\setS}_+^{n \times r}$, we have the error bound
	\beq \label{eq:jiang_bnd}
    \dist(\bX,\setS_+^{n,r}) \leq \sqrt{2r} \sqrt{\| \bX \bone \|_2^2 - r}.
    \eeq 
\end{Theorem}
The authors of \cite{jiang2023exact} actually derived a more general result than the above,
but the above result is considered the most representative and was the standard choice in the authors' numerical demonstrations.
From the perspective of this study, the most interesting question lies in how the above error bound was shown.
We will come back to this later. 
Using the error bound principle (applying \eqref{eq:jiang_bnd} to the penalized formulation \eqref{eq:cm_pen_part2}), the authors of  \cite{jiang2023exact} gave the following exact penalization formulation
\beq \label{eq:form_jiang}
\min_{\bX \in \tilde{\setS}_+^{n,r}} f(\bX) + \lambda \sqrt{\| \bX \bone \|_2^2 - r}
\eeq
for the ONMF problem \eqref{eq:prob_onmf} or for CM problems in the NSOM case.
The presence of a square root in the penalty of \eqref{eq:form_jiang} makes the objective function non-smooth, which adds some challenge from the viewpoint of building algorithms.
The authors of \cite{jiang2023exact} handled problem \eqref{eq:form_jiang} by using manifold optimization to deal with the manifold $\tilde{\setS}_+^{n \times r}$ and by applying smooth approximation to deal with the penalty term.
Taking inspiration from the above error bound result, another group of researchers derived a general error bound result as follows. 
\begin{Theorem}
	%[Theorem 2 in \cite{chen2022tight}] 
	\label{thm:chen}
	{\bf (Theorem {\revcolor5} in \cite{chen2022tight})}
	For any $\bX \in \Rbb^{n \times r}$, we have the error bound
	\beq \label{eq:chen_bnd}
	\dist(\bX, \setS_+^{n,r}) \leq 5 r^{\frac{3}{4}} ( \| \bX_- \|_{\rm F}^\half + \| \bX^\top \bX - \bI \|_{\rm F}^\half ),
	\eeq 
	where $\bX_- = \max\{ -\bX, \bzero \}$.
\end{Theorem}
%Here, $\max\{ \bA, \bB \}$ denotes the element-wise extension of $\max\{ a, b \}$.
%The authors of \cite{chen2022tight} also used analysis to argue that the order of the error bound, i.e., $1/2$, may not be made larger.
% The analyses in \cite{jiang2023exact,chen2022tight} also suggest that the order of the error bound functions in \eqref{eq:jiang_bnd} and \eqref{eq:chen_bnd}, i.e., $1/2$, may not be made larger.
% In the first part of this study, we used Theorem~\ref{thm:chen} to show an inexact penalization result for EXPP when the constraint set is replaced by the non-negative unit spectral norm ball $\setB^{n,r}_+$.

\subsection{A Modified Error Bound and a New EXPP for NSOMs}

%\subsection{A Modified Error Bound for Non-Negative Semi-Orthogonal Matrices}

We derive a modified error bound result for $\setS_+^{n,r}$. 
%under the theme of EXPP.
Let
\[
\tilde{\setB}_+^{n \times r}= \{ \bX \in \Rbb_+^{n \times r} \mid \| \bx_j \|_2 \leq 1 ~ \forall j \}
\]
define the non-negative column-wise unit $\ell_2$-norm ball.
Let
\beq \label{eq:penalty_psi_nsom}
\psi_p(\bX) = \left[ \sum_{j=1}^r c_1(\bx)^p + \sum_{i=1}^n \rho_1(\bar{\bx}_i)^p \right]^\frac{1}{p}
\eeq 
be a penalty function, where $p \in \{1,2\}$,
\[
c_1(\bx) = 1 - \| \bx \|_2, \quad 
\rho_1(\bx) = \| \bx \|_1 - \| \bx \|_\infty.
\]
The penalty $\psi_p$ shares a similar rationale as the one in \eqref{eq:penalty_ppm_interpret}:
%for $\bX \in \tilde{\setB}_+^{n \times r}$, 
For $\bX \in \Rbb^{n \times r}$ with $\| \bx_j \|_2 \leq 1$ for all $j$,
$c_1(\bx_j)$ promotes $\bx_j$ to have unit $\ell_2$ norm, while $\rho_1(\bar{\bx}_i)$ promotes $\bar{\bx}_i$ to be a scaled unit vector.
We should note that $\| \bx \|_1 \geq \| \bx \|_\infty$, and that $\| \bx \|_1 = \| \bx \|_\infty$ if and only if $\bx$ is a scaled unit vector.
We should also recall 
%that the characterization in 
from
\eqref{eq:nsom_char} 
%asserts 
that an NSOM has rows being scaled unit vectors.
Our result is as follows.
\begin{Theorem} \label{thm:eb2_Snr+}
	{\bf (error bound for non-negative semi-orthogonal matrices)}
	For any $\bX \in \tilde{\setB}^{n \times r}_+$, we have the error bounds 
        \begin{align}  \label{eq:eb2_Snr+}    
        \dist(\bX,\setS^{n,r}_+)
			 & \leq \nu \, \psi_2(\bX) \leq \nu \, \psi_1(\bX),
        \end{align} 
	where $\nu = \max\{ \sqrt{6}, 2 \sqrt{r} \}$. 
\end{Theorem}

The proof of Theorem~\ref{thm:eb2_Snr+} is provided in Appendix \ref{app:thm:eb2_Snr+}, and we will give insight into these error bounds in the next subsection.
Theorem~\ref{thm:eb2_Snr+} can be used to build a new EXPP formulation.
From \eqref{eq:eb2_Snr+} we have a further error bound 
\[
\nu \, \psi_1(\bX) \leq \nu \left[  r - \| \bX \|_\fro^2 + \sum_{i=1}^n ( \bone^\top \bar{\bx}_i - s_1(\bar{\bx}_i) ) \right], 
\]
where we use the fact that $\| \bx_j \|_2 \geq \| \bx_j \|_2^2$ whenever $\| \bx_j \|_2 \leq 1$. 
We recall $s_1(\bx) = x_{[1]}$;
$\bone^\top \bar{\bx}_i - s_1(\bar{\bx}_i)$ is an alternate form of $\rho_1(\bar{\bx}_i)$ for non-negative $\bar{\bx}_i$.
The above error bound leads us to the following new EXPP formulation for the NSOM case:
\beq \label{eq:new_expp_nsom}
\min_{\bX \in \tilde{\setB}_+^{n \times r}} F_\lambda(\bX)= f(\bX) + \lambda 
%[  \textstyle \sum_{i=1}^n ( \bone^\top \bar{\bx}_i - s_1(\bar{\bx}_i) ) - \| \bX \|_\fro^2 ].
\left[   \sum_{i=1}^n ( \bone^\top \bar{\bx}_i - s_1(\bar{\bx}_i) )  - \| \bX \|_\fro^2  \right].
\eeq
Unlike the previous EXPP formulations, 
%the formulation in \eqref{eq:new_expp_nsom} 
the above formulation
has non-smooth penalty terms $-s_1(\bar{\bx}_i)$'s.
However, since they are concave, they are considered not difficult to handle when we use majorization-minimization techniques to build algorithms.
Specifically, by Jensen's inequality, we have
\[
-s_1(\bx) \leq -s_1(\bx')+x_{l^\prime}^\prime-x_{l^\prime},
\quad \text{for any $\bx, \bx'$,}
\]
where $l'$ is such that $x'_{l'} = s_1(\bx')$.
The above inequality can be used to conveniently construct a majorant for the penalty function.
Moreover, it should be mentioned that the projection onto $\tilde{\setB}_+^{n \times r}$ has a closed form.
The projection $\Pi_{\tilde{\setB}_+^{n \times r}}(\bZ)$ for a given matrix $\bZ  \in \Rbb^{n \times r}$ corresponds to the projections $\Pi_{\setB_+^n}(\bz_j)$'s, where $\setB_+^n = \setB^n \cap \Rbb_+^n$.
The projection $\Pi_{\setB_+^n}(\bz)$ for a given vector $\bz \in \Rbb^n$ equals $\bz_+$ if $\| \bz_+ \|_2 \leq 1$, and $\bz_+/\| \bz_+ \|_2$ if $\| \bz_+ \|_2 > 1$;
here, $\bz_+ = \max \{ \bzero,\bz \}$.

\subsection{Insights and Further Remarks for NSOMs}

We should describe the insight behind the proof of Theorem \ref{thm:eb2_Snr+}.
Some of the key steps of our proof are actually the same as those of the previous results in Theorems~\ref{thm:jiang_bnd} and \ref{thm:chen}.
We however change one important ingredient.
Recall from \eqref{eq:nsom_char} that every NSOM has rows being scaled unit vectors.
This motivates us to consider the scaled unit vector set
\[
\setW^n := \{ \bx \in \Rbb^n \mid \bx = \alpha \be_i, ~ \alpha \in \Rbb, i \in \{1,\ldots,n \} \}.
\]
In particular we consider the error bound for $\setW^n$.
\begin{Lemma} \label{lem:setW}
    For any $\bx \in \Rbb^n$, we have the error bound
    \beq
        \dist(\bx,\setW^n) \leq \| \bx \|_1 - \| \bx \|_\infty = \rho_1(\bx).
    \eeq 
\end{Lemma}
\medskip
{\em Proof of Lemma~\ref{lem:setW}:} \
%Given an $\bx \in \Rbb^n$,
Let $\bx \in \Rbb^n$ be given.
Let $\by = x_l \be_l$, where $l$ is such that $|x_l| = \max\{ |x|_1,\ldots,|x_n| \}$.
It can be verified that $\by = \Pi_{\setW^n}(\bx)$ and  $\dist(\bx,\setW^n) = \| \bx - \by \|_2 \leq \| \bx - \by \|_1 = \rho_1(\bx)$.
\hfill $\blacksquare$
\medskip

\noindent
Our proof differs in that we use the error bound in Lemma~\ref{lem:setW}, while the prior results %did not use the error bound notion and derived a different bound.
used a different bound.
Interested readers are referred to Appendix \ref{app:diff_onmf_er}, wherein we delineate the difference.

In fact, from the proof of Theorem \ref{thm:eb2_Snr+}, we notice the following generalization.
\begin{Corollary}
    Let $\setS^{n,r}_{\sf e} = \{ \bX \in \Rbb^{n \times r} \mid \| \bx_j \|_2 = 1, \, \bar{\bx}_i \in \setW^r, ~ \forall i,j \}$. For any $\bX \in \Rbb^{n \times r}$ with $\| \bx_j \|_2 \leq 1$ for all $j$, we have the error bounds $\dist(\bX,\setS^{n,r}_{\sf e}) \leq \nu \, \psi_2(\bX) \leq \nu \, \psi_1(\bX)$,
    where $\nu$ is given by the one in Theorem \ref{thm:eb2_Snr+}.
\end{Corollary}
We omit the proof as it is a trivial variation of the proof of Theorem \ref{thm:eb2_Snr+}.
Note that $\setS^{n,r}_+ \subset \setS^{n,r}_{\sf e} \subset \setS^{n,r}$,
and $\setS^{n,r}_{\sf e}$ permits negative components.
%Note that an $\bX \in \setS^{n,r}_{\sf e}$ can have both negative and non-negative components,
%and if we restrict $\bX$ to have non-negative components then $\bX$ lies in $\setS_+^{n,r}$.

It is also worth noting that
Theorem \ref{thm:eb2_Snr+} has connections with the previous results in Theorems~\ref{thm:jiang_bnd} and \ref{thm:chen}.
%may be used to generate the previous error bounds in Theorems~\ref{thm:jiang_bnd} and \ref{thm:chen}.
%,  subject to differences with the constant $\nu$ (which may be unimportant).
To describe it, we first derive the following result.
\begin{Lemma} \label{eq:lem_norm_diff}
	For any $\bx \in \Rbb^n$, 
    %we have $( \| \bx \|_1 - \| \bx \|_\infty )^2 \leq \| \bx \|_1^2 - \| \bx \|_2^2$.
    it holds that $\rho_1(\bx)^2 \leq \| \bx \|_1^2 - \| \bx \|_2^2 = \rho_2(\bx)$.
\end{Lemma}
\medskip
{\em Proof of Lemma \ref{eq:lem_norm_diff}:} \
Without loss of generality, assume that $| x_1 | \geq | x_i |$ for all $i$. We have
\begin{align*}
	\rho_1(\bx)^2  & =
	( \| \bx \|_1 - |x_1| )^2 = \| \bx \|_1^2 - 2 |x_1| \| \bx \|_1 + |x_1|^2  \\
     & \leq \| \bx \|_1^2 - 2 (|x_1|^2 + |x_2|^2 + \cdots + |x_n|^2) + |x_1|^2 \\
     & \leq \| \bx \|_1^2 - \| \bx \|_2^2.
 \end{align*}
% \begin{align*}
% 	\rho_1(\bx)^2  & =
% 	\textstyle ( \sum_{i=2}^n |x_i| )^2  \\
% 	 & \textstyle =\sum_{i=2}^n \sum_{j=2}^n |x_i| |x_j| \\
% 	 & \textstyle =\sum_{i=1}^n \sum_{j=1}^n |x_i| |x_j| - |x_1|^2 - 2 |x_1| \sum_{i=2}^n |x_i| \\
% 	 & \textstyle \leq \sum_{i=1}^n \sum_{j=1}^n |x_i| |x_j| - |x_1|^2 - |x_1| \sum_{i=2}^n |x_i| \\ 
% 	 &  \textstyle \leq \sum_{i=1}^n \sum_{j=1}^n |x_i| |x_j| - |x_1|^2 -  \sum_{i=2}^n |x_i|^2 \\
% 	 & = \| \bx \|_1^2 - \| \bx \|_2^2.
%  \end{align*}
% \begin{align*}
% 	( \| \bx \|_1 - \| \bx \|_\infty )^2 & =
% 	\textstyle ( \sum_{i=2}^n |x_i| )^2  \\
% 	 & \textstyle =\sum_{i=2}^n \sum_{j=2}^n |x_i| |x_j| \\
% 	 & \textstyle =\sum_{i=1}^n \sum_{j=1}^n |x_i| |x_j| - |x_1|^2 - 2 |x_1| \sum_{i=2}^n |x_i| \\
% 	 & \textstyle \leq \sum_{i=1}^n \sum_{j=1}^n |x_i| |x_j| - |x_1|^2 - |x_1| \sum_{i=2}^n |x_i| \\ 
% 	 &  \textstyle \leq \sum_{i=1}^n \sum_{j=1}^n |x_i| |x_j| - |x_1|^2 -  \sum_{i=2}^n |x_i|^2 \\
% 	 & = \| \bx \|_1^2 - \| \bx \|_2^2.
%  \end{align*}
The proof is complete.
\hfill $\blacksquare$
\medskip

\noindent From Lemma~\ref{eq:lem_norm_diff} we further have the following result.
%\begin{Corollary}   \label{cor:eb2_Snr+_1}
\begin{Lemma}   \label{cor:eb2_Snr+_1}
    For any $\bX \in \tilde{\setB}^{n \times r}_+$, 
    % we have the error bound
    it holds that
    \begin{subequations} \label{eq:eb2_Snr+_altalt}
        \begin{align}
            %\dist(\bX,\setS^{n,r}_+)
            \psi_2(\bX) 
			 & \leq  \sqrt{r + \| \bX \bone \|_2^2 - 2 \| \bX \|_\fro^2} \label{eq:eb2_Snr+_alt_a} \\
             &  =  \| \bX^\top \bX - \bI \|_{\ell_1}^\half
             \label{eq:eb2_Snr+_alt_b} \\
             & \leq r \| \bX^\top \bX - \bI \|_{\fro}^\half.
             \label{eq:eb2_Snr+_alt_c}
        \end{align}    
    \end{subequations}
%    Consequently, $\sqrt{r + \| \bX \bone \|_2^2 - 2 \| \bX \|_\fro^2}$ and $\| \bX^\top \bX - \bI \|_{\ell_1}^\half$ are effectively error bound functions for $\setS^{n,r}_+$ relative to $\tilde{\setB}^{n \times r}_+$.
    %where $\nu= ???$.
%\end{Corollary}
\end{Lemma}
\medskip
{\em Proof of 
%Corollary 
Lemma
\ref{cor:eb2_Snr+_1}:} \
Let $\bX \in \tilde{\setB}^{n \times r}_+$ be given.
Since $\| \bx_j \|_2 \leq 1$,
we have $c_1(\bx_j)^2 \leq 1 - \| \bx_j \|_2 \leq 1 - \| \bx_j \|_2^2 = c_2(\bx_j)$.
By Lemma~\ref{eq:lem_norm_diff} we have $\rho_1(\bar{\bx}_i)^2 \leq \rho_2(\bar{\bx}_i)$.
Applying these results to \eqref{eq:penalty_psi_nsom} and observing \eqref{eq:penalty_ppm_interpret}, we have \eqref{eq:eb2_Snr+_alt_a}.
Eq.~\eqref{eq:eb2_Snr+_alt_b} is a direct consequence of  Lemma~\ref{lem:q_identity}.
Eq.~\eqref{eq:eb2_Snr+_alt_c} is the consequence of the norm result $\| \ba \|_1 \leq \sqrt{n} \| \ba \|_2$ for $\ba \in \Rbb^n$.
\hfill $\blacksquare$
\medskip

\noindent 
%From \eqref{eq:eb2_Snr+_altalt} we see the following.
By Theorem~\ref{thm:eb2_Snr+}, all the functions in \eqref{eq:eb2_Snr+_altalt} are effective error bound functions of $\setS_+^{n,r}$ relative to $\tilde{\setB}_+^{n \times r}$.
If we 
%restrict $\tilde{\setB}_+^{n \times r}$ to $\tilde{\setS}_+^{n \times r}$, 
limit $\bX$ to lie in $\tilde{\setS}_+^{n \times r}$,
then the effective error bound function in \eqref{eq:eb2_Snr+_alt_a} becomes that in \eqref{eq:jiang_bnd} in Theorem~\ref{thm:jiang_bnd}. 
The effective error bound function in \eqref{eq:eb2_Snr+_alt_c} is identical to that in \eqref{eq:chen_bnd} in Theorem~\ref{thm:chen} for the special case of $\bX \in \tilde{\setB}_+^{n \times r}$.
Hence, simply speaking, Theorem~\ref{thm:eb2_Snr+} can produce the error bound results in the prior studies.

We end with a further comment.
The penalty functions $\rho_1$ and $\rho_2$ for scaled unit vectors were considered in \cite{wang2021clustering}.
The authors of that work studied a variation of the ONMF problem \eqref{eq:prob_onmf0}, wherein they remove the unit $\ell_2$-norm column constraints with $\bX$.
They showed stationarity results with the use of $\rho_1$ and $\rho_2$.
Their analysis is not based on 
%the notion of 
error bounds, 
%, and thus it is different from our analysis.
%We should also point out that the inclusion of the unit $\ell_2$-norm column constraints makes the problem nature rather different.
%Also, 
and the problem nature with $\setS_+^{n,r}$ is considered more challenging than that with the scaled unit vectors.

\section{Numerical Results in Different Applications}
\label{sect:p2_num}

In this section we numerically demonstrate EXPP in different applications.

\subsection{The Algorithm for EXPP}

The algorithm used to implement EXPP in our experiments is an extrapolated variant of the projected gradient (PG) method, which was numerically found to have faster convergence; see, e.g., \cite{xu2013block}.
%as reported in the literature {\red [X]. (I will do it)}
It is also the same algorithm used in our prior studies \cite{shao2019framework,shao2020binary}.
The algorithm is shown in Algorithm \ref{alg:hot}.
To explain it, let us 
write down the EXPP formulation 
\beq \label{eq:alg_obj}
\min_{\bx \in \setX} F_\lambda(\bx) =f(\bx)+\lambda \, h(\bx),
\eeq 
where $\setX = \conv(\setV)$; $h(\bx)= - \| \bx \|_2^2$; $f$ is assumed to be differentiable and have Lipschitz continuous gradient on $\setX$.
Lines 6--7 of Algorithm \ref{alg:hot} constitute the extrapolated PG step.
If we replace $\bz$ by $\tilde{\bx}^l$ and $\nabla G_{\lambda_k}( \bz | \tilde{\bx}^{l} )$ by $\nabla F_{\lambda_k}(\tilde{\bx}^l)$, we return to the baseline PG step.
The point $\bz$ is an extrapolated point, and the extrapolation sequence $\{ \alpha_l \}$ is chosen as the FISTA sequence \cite{beck2017first}.
The function $G_\lambda(\bx|\bx')$ is a majorant of $F_\lambda(\bx)$ at $\bx'$, and we apply majorization before the PG step.
%Specifically we majorize some concave term in $F_\lambda(\bx)$, namely, 
%We majorize some concave term in $F_\lambda(\bx)$.
%, as an optional operation.
Specifically we majorize the concave $h$ by 
%$u(\bx|\bx') =  \| \bx ' \|_2^2 - 2 \bx^\top \bx'$ 
 $u(\bx|\bx') =  -\| \bx ' \|_2^2 - 2 (\bx-\bx')^\top \bx'$
(it is the result of Jensen's inequality),
and then we construct the majorant $G_\lambda(\bx|\bx')= {\revcolor f(\bx)} + \lambda u(\bx|\bx')$ (for $\lambda > 0$).
With this choice we set the step size $\eta$ as the reciprocal of the Lipschitz constant of $\nabla G_\lambda(\bx|\bx')$.
Furthermore, 
%it should be noted that 
we apply a homotopy strategy, wherein we gradually increase the penalty parameter $\lambda$ so that we start with a possibly convex problem and end with the target EXPP problem (with exact penalization).

\begin{algorithm}[!hbt]
	%\caption{An PGD type of algorithm following homotopy optimization priciple.} 
    \caption{A PG-type algorithm for \eqref{eq:alg_obj}, with homotopy}
    \label{alg:hot}
	\begin{algorithmic}[1]
		
		\STATE \textbf{given:} a sequence $\{ \lambda_k \}$, an extrapolation sequence $\{\alpha_l\}$, and a starting point $\bx^0$
		\STATE $k \leftarrow 0$
		
		\REPEAT 
		  \STATE   
            %$\tilde{\bx}^0=\bx^k$, $\bz=\bx^k$,
            $\tilde{\bx}^0  \leftarrow \bx^k$, $\bz \leftarrow \bx^k$,
            %$\tilde{\bx}^0 \leftarrow \bx^k$, $\bz \leftarrow\bx^k$,
            $l\leftarrow 0$
            \REPEAT
            \STATE $\tilde{\bx}^{l+1}  \leftarrow \Pi_{\setX}( \bz - \eta \nabla G_{\lambda_k}( \bz | \tilde{\bx}^{l} ) )$ for some $\eta > 0$
            %\STATE $\tilde{\bx}^{l+1}=\Pi_{\setX}\left(\bz -\eta_k\nabla_{\boldsymbol{z}} G(\boldsymbol{z}\mid\tilde{\bx}^{l} )\right)$
            \STATE %$\bz=\tilde{\bx}^{l+1}+\alpha_l(\tilde{\bx}^{l+1}-\tilde{\bx}^{l})$\
            $\bz \leftarrow \tilde{\bx}^{l+1} + \alpha_l(\tilde{\bx}^{l+1}-\tilde{\bx}^{l})$
            \STATE
            $l \leftarrow l+1$
            \UNTIL{a stopping rule is met}
		\STATE $\bx^{k+1}\leftarrow\tilde{\bx}^{l}$
		\STATE $k \leftarrow k+1$
		
		\UNTIL{a stopping rule is met}
		
		\STATE \textbf{output:} $\bx^k$

	\end{algorithmic}
\end{algorithm}
Algorithm \ref{alg:hot} is also used to implement the new EXPP formulations \eqref{eq:new_expp_ppm}, \eqref{eq:new_expp_sam}, and \eqref{eq:new_expp_nsom} for the PPM, SAM, and NSOM cases, respectively.
We will call them ``EXPP--II'' to avoid confusion with the previous EXPP.
For the PPM and SAM cases, we have $h(\bX) = \| \bX \bone \|_2^2 - 2 \| \bX \|_\fro^2$.
We choose $u(\bX|\bX') = \| \bX \bone \|_2^2 - 4 \tr(\bX^\top \bX') + 2 \| \bX' \|_\fro^2$; we keep the convex term $\| \bX \bone \|_2^2$ and majorize the concave term $-2 \| \bX \|_\fro^2$.
For the NSOM case, we have 
$h(\bX) = -\| \bX \|_{\fro}^2 + \sum_{i=1}^n (\bone^\top \bar{\bx}_i - s_1(\bar{\bx}_i))$.
We choose $u(\bX|\bX') = - 2\tr(\bX^\top \bX') + \| \bX' \|_\fro^2 + \sum_{i=1}^n (\bone^\top \bar{\bx}_i - x_{i,l_i'})$,
where $l_i'$ is such that $x_{i,l_i'}'= \max \{ x_{i,1}',\ldots,x_{i,r}' \}$;
again we keep the convex terms and majorize the concave terms.
The rest of the operations are identical.

{\revR1color We should mention that Algorithm \ref{alg:hot} is equipped with guarantees of convergence to a stationary point.
To be specific, the majorization and extrapolated PG loop in Lines 5--9 of Algorithm \ref{alg:hot} were shown to be able to converge to a stationary point of problem \eqref{eq:alg_obj} (with $\lambda = \lambda_k$) \cite{shao2019framework}.
The premise of this is that $h$ is differentiable and has Lipschitz continuous gradient.
This premise is satisfied in most of the above described cases, 
with the NSOM case being the only exception.
The critical-point convergence for the NSOM case (nonsmooth $h$) is a subject of future study.
}

We may deal with applications that have non-differentiable objective function $f$. 
For such cases we replace the extrapolated PG step in Lines 6--7 of Algorithm \ref{alg:hot} with the projected subgradient method.
%; we use the non-summable diminishing step size rule.

Some details with the stopping rules of Algorithm \ref{alg:hot} are as follows.
Unless specified otherwise,
the stopping rule for the inner loop is either $\| \tilde{\bx}^{l+1} - \tilde{\bx}^l \|_2/ \| \tilde{\bx}^l \|_2 < \varepsilon_1$ or $l > \bar{L}$ for some given $\varepsilon_1$ and $\bar{L}$.
Unless specified otherwise,
the stopping rule for the outer loop is $\| \bx^{k+1} - \bx^k \|_2/ \| \bx^k \|_2 < \varepsilon_2$, $\dist(\bx^k,\setV) \leq \varepsilon_3$, or $\lambda_k > \bar{\lambda}$ for some given $\varepsilon_2$, $\varepsilon_3$, and $\bar{\lambda}$.

{\revMovecolor\subsection{MIMO Detection with MPSK Constellations}

We consider MIMO detection.
In this problem, we have a received signal model $\by = \bH \bx + \bv$, where $\by \in \Cbb^m$ is the received signal; $\bH \in \Cbb^{m \times n}$ is the channel; $\bx \in \Cbb^n$ is the transmitted symbol vector; $\bv \in \Cbb^m$ is noise.
The goal is to detect $\bx$ from $\by$.
Assuming that every $x_i$ is drawn from an $M$-ary phase shift keying (PSK) constellation $\Theta_M = \{ x \in \Cbb \mid x= e^{\jj \frac{2\pi l}{M} + \jj \frac{\pi}{M} }, ~ l \in \{0,1,\ldots,M-1\} \}$, we consider the maximum-likelihood (ML) detector
\beq 
\hat{\bx} = \arg \min_{\bx \in \Theta_M^n} \| \by - \bH \bx \|_2^2. \nonumber 
\eeq 
We use EXPP to handle the ML detector.

The benchmarked algorithms are
(i) MMSE: the minimum mean square detector;
(ii) SDR: the semidefinite relaxation detector \cite{ma2004semidefinite};
%(with the number of randomization equal to $1,000$
(iii) LAMA: the approximate message passing method in \cite{jeon2018optimal} with incorporation of damping \cite{rangan2019convergence} (with damping factor $0.7$).
The parameter settings of EXPP are $\lambda_0 = 0.01$, $\lambda_{k+1} = 5 \lambda_k$, $\varepsilon_1 = 10^{-4}$, $\bar{L}= 100$,  $\bar{\lambda} = 10^4$.

The simulation settings are as follows.
A number of $10,000$ Monte Carlo trials were run.
The channel $\bH$ was generated based on a correlated MIMO channel model $\bH = \bR^\half_r \tilde{\bH} \bR_t^\half$,
where the components of $\tilde{\bH}$ are independent and identically distributed (i.i.d.) and follow a circular complex Gaussian distribution with mean zero and variance $1$,
and $\bR_t$ and $\bR_r$ follow the exponential model with parameter $\rho= 0.2$ \cite{loyka2001channel}.
The vector $\bv$ was generated as a component-wise i.i.d. circular complex Gaussian noise with mean $0$ and variance $\sigma^2$.

\ifconfver
\begin{figure} [htb!]
    \centering
\includegraphics[scale=0.48]{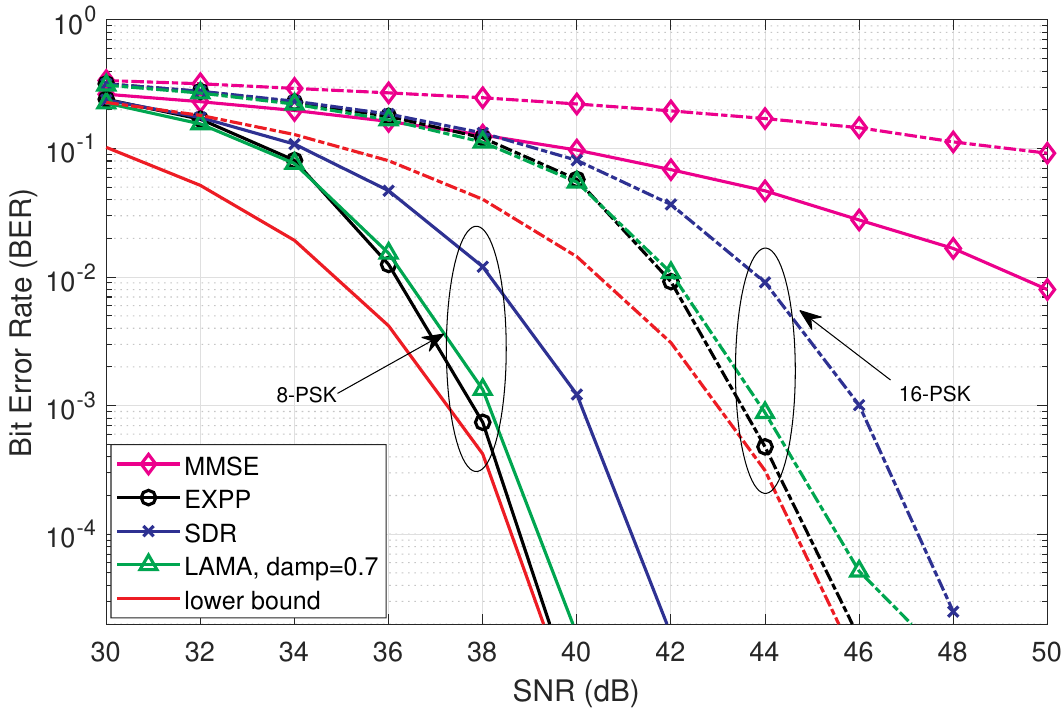}
    \caption{{\revMovecolor MIMO detection BER performance. Solid lines: $8$-PSK; dashed lines: $16$-PSK.}}
    \label{fig:mimo_det}
\end{figure}
\else
\begin{figure} [htb!]
    \centering
\includegraphics[scale=0.6]{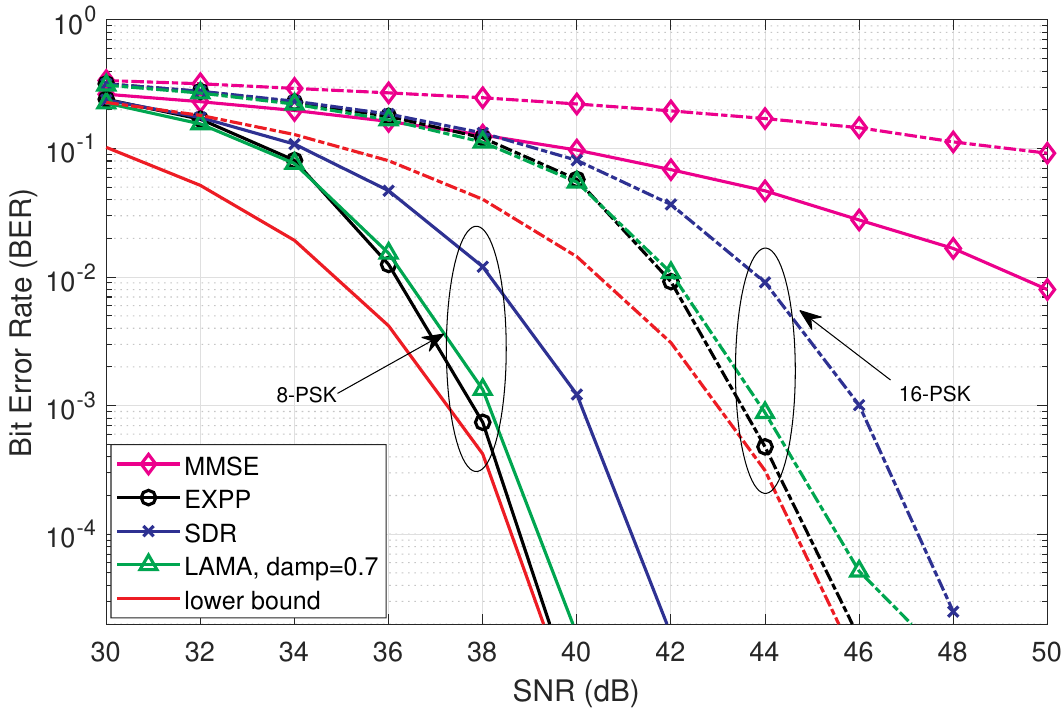}
    \caption{{\revMovecolor MIMO detection BER performance. Solid lines: $8$-PSK; dashed lines: $16$-PSK.}}
    \label{fig:mimo_det}
\end{figure}
\fi

\begin{table}[htb!]
\centering
\scalebox{1}{
    \centering
\begin{tabular}{c|c|c|c}
\hline Algorithms & SDR & EXPP & LAMA \\
\hline Runtime (in sec.) & 1.017 & 0.026 & 0.021 \\ \hline
\end{tabular}}
    \caption{{\revMovecolor Average runtime performance in MIMO detection. $8$-PSK, SNR$=38$dB.}}
    \label{tab:mimo}
\end{table}

Fig.~\ref{fig:mimo_det} and Table~\ref{tab:mimo} display the bit-error-rate (BER) and runtime performances of the various detectors, respectively.
The problem size is $m=n=80$.
``Lower bound'' is the performance baseline when there is no MIMO interference.
EXPP gives the best BER performance and is closely followed by LAMA.
The runtime performance of EXPP is slightly worse than that of LAMA but still competitive.}

% DKS figures here seems more nice.
\begin{figure*}[htb]
     \centering

     \begin{subfigure}[htb]{0.32\textwidth}
         \centering
         \includegraphics[width=\textwidth]{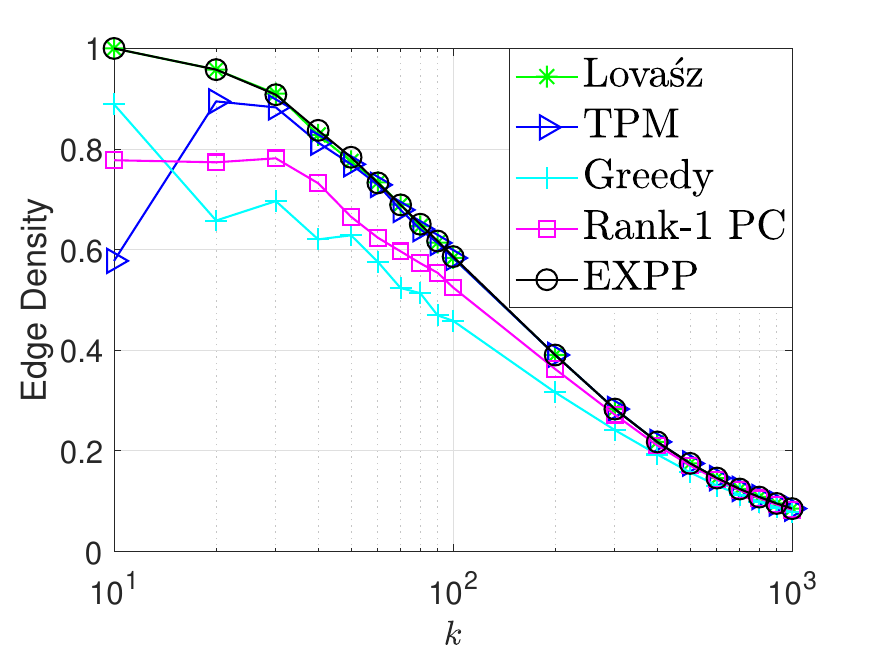}
         \caption{Loc-Gowalla}
     \end{subfigure}
           \hfill
        \begin{subfigure}[htb]{0.32\textwidth}
         \centering
         \includegraphics[width=\textwidth]{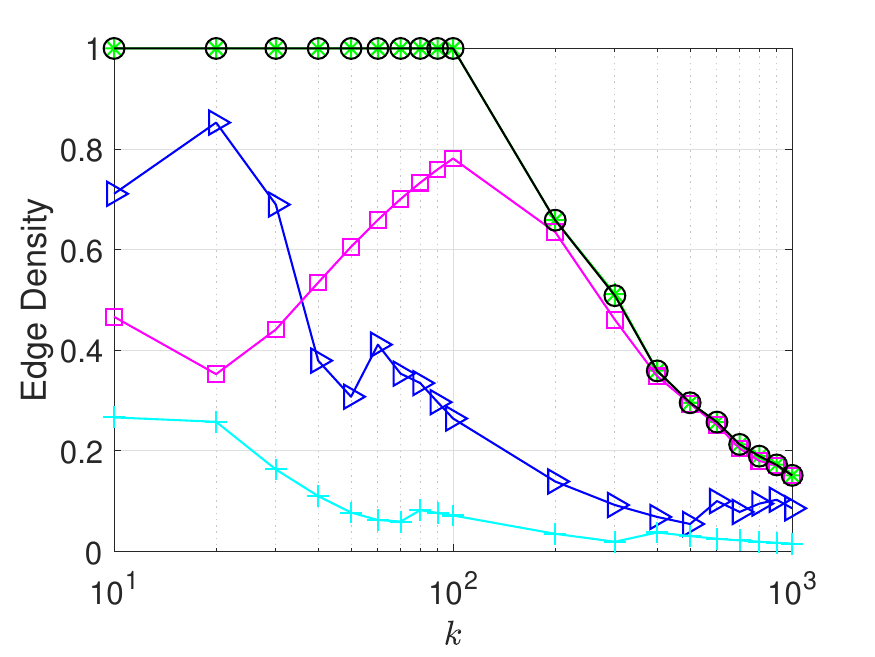}
         \caption{{\revcolor NotreDame}}
     \end{subfigure}
          \hfill
        \begin{subfigure}[htb]{0.32\textwidth}
         \centering
         \includegraphics[width=\textwidth]{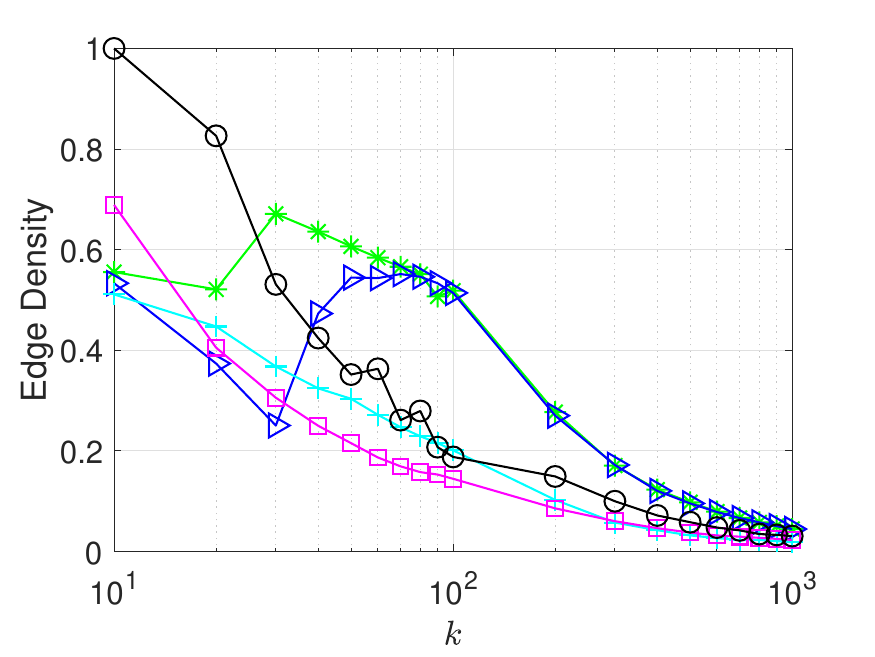}
         \caption{Google}
     \end{subfigure}
     \hfill
        \begin{subfigure}[htb]{0.32\textwidth}
         \centering
         \includegraphics[width=\textwidth]{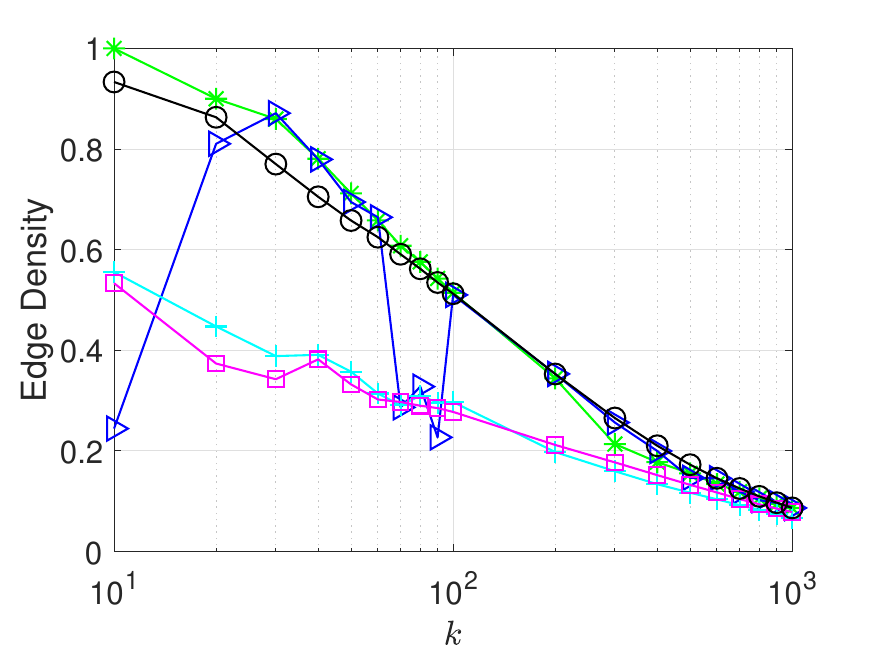}
         \caption{YouTube}
     \end{subfigure}
     \hfill
     \begin{subfigure}[htb]{0.32\textwidth}
         \centering
         \includegraphics[width=\textwidth]{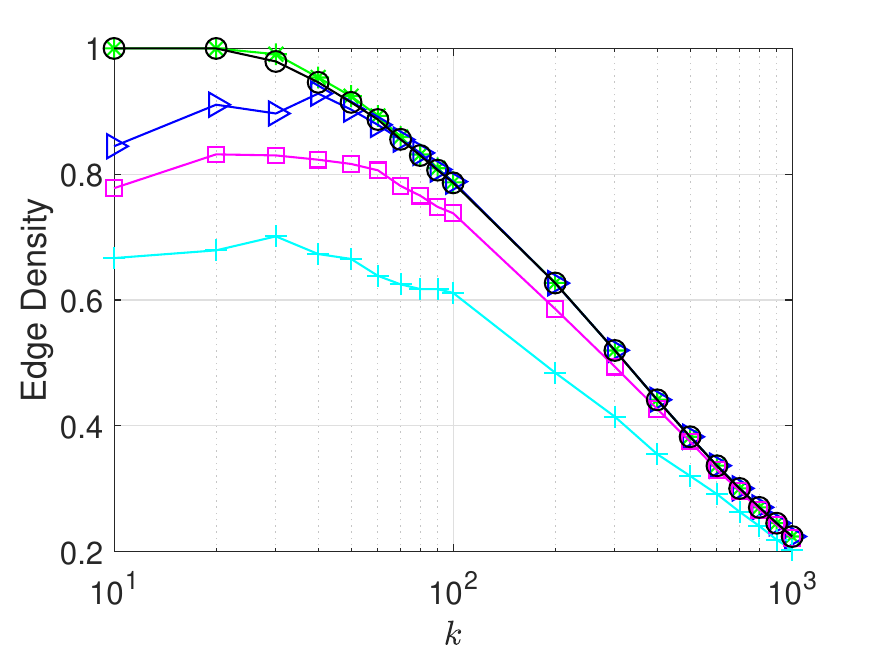}
         \caption{Talk}
    \end{subfigure}
     %  \hfill
     % \begin{subfigure}[htb]{0.32\textwidth}
     %     \centering
     %     \includegraphics[width=\textwidth]{figs/DKS/DKS_roadnet.eps}
     %     \caption{RoadNet}
     % \end{subfigure}
    \hfill
     \begin{subfigure}[htb]{0.32\textwidth}
         \centering
         \includegraphics[width=\textwidth]{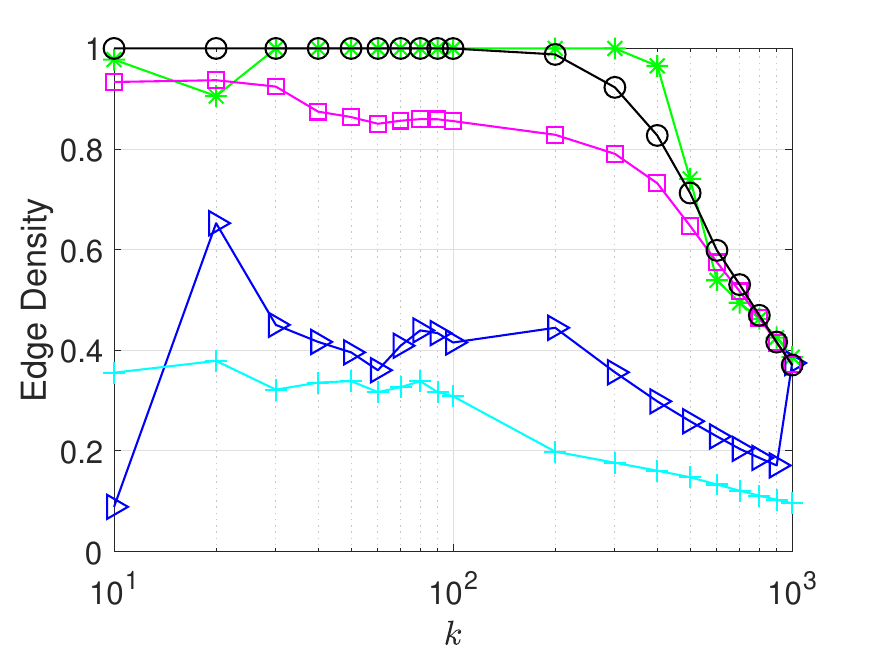}
         \caption{LiveJournal}
     \end{subfigure}
\caption{Performance with the D$k$S problem.}
\label{fig:dks_exp_sup}
\end{figure*}

\subsection{Densest $k$-Subgraph Problem}
We consider the densest $k$-subgraph  (D$k$S) problem.
The problem is to identify the most connected subgraph from a graph.
Let $\bW \in \Rbb^{n \times n}$ be the adjacency matrix of a given graph.
Given a subgraph size $k$,
the D$k$S problem is given by
\[
\min_{\bx \in \setU_k^n} - \bx^\top \bW \bx, 
\]
where $\setU_k^n = \{ \bx \in \{ 0, 1 \}^n \mid \bone^\top \bx = k \}$.
We use EXPP to handle this problem.

The benchmarked algorithms are
(i) Greedy: the greedy algorithm in \cite{feige2001dense};
(ii) TPM: the truncated power method \cite{yuan2013truncated};
(iii) Rank-$1$ PC: the rank-one binary principal component approximation method \cite{papailiopoulos2014finding};
(iv) Lova\'{s}z:  Lova\'{s}z relaxation, implemented by the ADMM method and a Frank-Wolfe post-processing step \cite{konar2021exploring}. 
The parameter settings of EXPP are $\lambda_0 = - \sigma_1(\bW)$, $\lambda_{k+1} = \lambda_k + 0.2 \sigma_1(\bW)$, $\varepsilon_1 = 10^{-3}$, $\bar{L}= 50$, $\varepsilon_3 = 10^{-2}$,  $\bar{\lambda} = 1.1\sigma_1(\bW)$.
Note that the above choice of $\lambda_0$ is to make the corresponding EXPP problem convex.
If $\lambda_k < 0$, then we choose $u(\bx|\bx')= h(\bx)$ (no majorization).
%We majorize $-\lambda\|\boldsymbol{x}\|_2^2$ if $\lambda\geq0$.

We performed our test on large real-world graphs \cite{leskovec2014snap}.
Table \ref{tab:dks_rt} describes the graph sizes of the tested datasets, which range from nearly $200,000$ nodes to {\revcolor nearly} $4,000,000$ nodes.
%with millions of nodes and edges, as listed in Table \ref{tab:dks_rt}.
We adopt the same data pre-processing as that in \cite{konar2021exploring}.
Fig.~\ref{fig:dks_exp_sup} shows the edge densities, $\bx^\top \bW \bx/(k^2 - k)$, achieved by  the various algorithms.
Lova\'{s}z gives the best performance in all the datasets.
EXPP is the second best.
Except for the Google dataset, EXPP achieves similar performance as Lova\'{s}z.
Table \ref{tab:dks_rt} shows the runtimes.
We see that EXPP runs faster than Lova\'{s}z.
%on the tested datasets.

\ifconfver
\begin{table*}[ht]
	\centering
 \scalebox{1}{
	\begin{tabular}{c|c|c||c|c|c|c|c}
		\hline
		\cline{1-6}
		\multirow{2}{*}{Datasets} &  \multirow{2}{*}{$n$} & \multirow{2}{*}{$m$} & \multicolumn{5}{|c}{Runtime (in sec.)} \\ \cline{4-8}
		&&& Lova\'{s}z & TPM & Greedy & Rank-1 PC & EXPP \\ \hline\hline
		Lcc-Gowalla & 196,591 &  950,327 & 41.32 & 0.10 & 0.02 & 0.56 & 2.40 \\ \hline
		{\revcolor NotreDame} & 325,729 & 1,497,134 & 18.95 & 0.07 & 0.03 & 0.91 & 3.11 \\ \hline
		Google& 875,713 & 5,105,039 & 274.13 & 0.50 & 0.13 & 5.9{\revcolor1} &  25.36\\ \hline
		YouTube & 1,134,890 & 2,987,624 & 1.05e3 & 1.04 & 0.13 & 2.94 & 18.27 \\ \hline
		Talk & 2,394,385 & 5,021,410 & 1.21e3 & 1.50 & 0.33 & 5.02 & 36.92 \\ \hline
		LiveJournal & {\revcolor3,997,962}  & {\revcolor34,681,189}  & 1.29e3 & 3.82 & 0.66 & 22.78 & 124.68 \\ \hline
	\end{tabular}}
	\caption{Average runtime with the D$k$S problem. $n = $ number of nodes, $m = $ number of edges.}
	\label{tab:dks_rt}
\end{table*}
\else
\begin{table*}[ht]
	\centering
 \scalebox{0.9}{
	\begin{tabular}{c|c|c||c|c|c|c|c}
		\hline
		\cline{1-6}
		\multirow{2}{*}{Datasets} &  \multirow{2}{*}{$n$} & \multirow{2}{*}{$m$} & \multicolumn{5}{|c}{Runtime (in sec.)} \\ \cline{4-8}
		&&& Lova\'{s}z & TPM & Greedy & Rank-1 PC & EXPP \\ \hline\hline
		Lcc-Gowalla & 196,591 &  950,327 & 41.32 & 0.10 & 0.02 & 0.56 & 2.40 \\ \hline
		{\revcolor NotreDame} & 325,729 & 1,497,134 & 18.95 & 0.07 & 0.03 & 0.91 & 3.11 \\ \hline
		Google& 875,713 & 5,105,039 & 274.13 & 0.50 & 0.13 & 5.9{\revcolor1} &  25.36\\ \hline
		YouTube & 1,134,890 & 2,987,624 & 1.05e3 & 1.04 & 0.13 & 2.94 & 18.27 \\ \hline
		Talk & 2,394,385 & 5,021,410 & 1.21e3 & 1.50 & 0.33 & 5.02 & 36.92 \\ \hline
		LiveJournal & {\revcolor3,997,962}  & {\revcolor34,681,189}  & 1.29e3 & 3.82 & 0.66 & 22.78 & 124.68 \\ \hline
	\end{tabular}}
	\caption{Average runtime with the D$k$S problem. $n = $ number of nodes, $m = $ number of edges.}
	\label{tab:dks_rt}
\end{table*}
\fi

\subsection{Graph Matching}
\label{section:graph_match}

We consider the GM problem described in Section~\ref{sect:exa_gm}.
The benchmarked algorithms are
(i) PATH \cite{zaslavskiy2008path};
(ii) GNCCP \cite{liu2013gnccp};
(iii) LAGSA \cite{xia2010efficient}.
PATH was concisely reviewed in Section~\ref{sect:exa_gm}, and GNCCP and LAGSA are related methods.
% These algorithms all use the Frank-Wolfe method to handle related GM formulations. 
% We did not test EXPP as it is expensive to run on large graphs.
We employ EXPP--II in \eqref{eq:new_expp_ppm}.
The parameter settings of EXPP are $\lambda_0 = 10^{-5}$, $\lambda_{k+1} = 4\lambda_{k}$, $\varepsilon_1 = 10^{-4}$, $\varepsilon_2 = 10^{-5}$, $\bar{L}= 100$, $\bar{\lambda} = 10^{3}$.

We performed our test on graphs constructed from various image datasets.
We follow a standard protocol to extract graphs from a pair of images; see e.g., \cite{zhou2015factorized}.
We tested several image sets from $5$ datasets: CMU-House \cite{CMUhouse}, PASCAL \cite{leordeanu2012unsupervised}, 
DTU\cite{aanaes2016large}, DAISY \cite{tola2009daisy}, and SUIRD \cite{liu2020rotation}.
CMU-House and PASCAL are small graphs and are commonly used in GM papers; DTU, DAISY, and SUIRD are large graphs.
For each pair of images, we performed $10$ independent graph constructions and used them for experiments.

Table \ref{tab:GM_real} shows the average matching accuracies and runtimes of the different algorithms.
The matching accuracy is defined as the ratio of the number of correctly matched nodes to the node size $n$.
We did not test PATH and LAGSA on large-size graphs because they run too slowly.
EXPP--II gives the best matching accuracy performance for all the datasets.
More importantly, EXPP--II runs much faster than the other algorithms.
This is particularly so for large-size graphs.
\ifconfver
\begin{table}[htb!]
	\centering
 \scalebox{0.82}{
	\begin{tabular}{c|c||c|c|c|c|c}
		\hline
		Data & $n$ & Metric & EXPP--II & GNCCP & PATH & LAGSA \\ \hline \hline
		\multirow{2}{*}{CMU-House} & \multirow{2}{*}{20} & acc.     & \textbf{77.00} & 74.00 & 54.50 & 43.00 \\ \cline{3-7}
		&  & time   & \textbf{0.03} & 0.09 & 1.21 & 0.41 \\ \hline
		\multirow{2}{*}{Car in PASCAL} &  \multirow{2}{*}{20} & acc. & \textbf{82.50} & 65.00 & 59.50 & 61.00 \\ \cline{3-7}
		&  & time   & \textbf{0.02} & 0.04 & 1.22 & 0.76 \\ \hline
		\multirow{2}{*}{Motorbike in PASCAL}  &  \multirow{2}{*}{20}  & acc. & \textbf{96.00} & 80.50 & 70.00 & 73.50 \\\cline{3-7}
		&  & time   & \textbf{0.02} & 0.08 & 1.16 & 0.64 \\ \hline\hline
		\multirow{2}{*}{DTU-House}  &  \multirow{2}{*}{500}  & acc. & \textbf{83.62} & 60.64 & - & - \\\cline{3-7}
		&  & time   & \textbf{16} & 711 & - & -    \\ \hline
		\multirow{2}{*}{HerzJesu in DAISY}  &  \multirow{2}{*}{500} & acc.     & \textbf{82.56} & 49.18 & - & - \\\cline{3-7}
		&  & time   & \textbf{11} & 786 & - & -   \\ \hline
		\multirow{2}{*}{Semper in DAISY}  &  \multirow{2}{*}{500} & acc.     & \textbf{92.04} & 54.46 & - & -   \\\cline{3-7}
		&  & time   & \textbf{13} & 619 & - & -   \\ \hline
		\multirow{2}{*}{Stadium in SUIRD}  &  \multirow{2}{*}{500} & acc.    & \textbf{91.68} & 40.46 & - & -   \ \\ \cline{3-7}
		&  & time   & \textbf{10} & 435 & - & -     \\ \hline
	\end{tabular}}
	\caption{Graph matching results. 
    $n =$ number of nodes,
    %Performance with graph matching in real images. 
    acc. $=$ matching accuracy in percent, time $=$ runtime in seconds. }
	\label{tab:GM_real}
\end{table}
\else
\begin{table}[htb!]
	\centering
 \scalebox{0.85}{
	\begin{tabular}{c|c||c|c|c|c|c}
		\hline
		Data & $n$ & Metric & EXPP--II & GNCCP & PATH & LAGSA \\ \hline \hline
		\multirow{2}{*}{CMU-House} & \multirow{2}{*}{20} & acc.     & \textbf{77.00} & 74.00 & 54.50 & 43.00 \\ \cline{3-7}
		&  & time   & \textbf{0.03} & 0.09 & 1.21 & 0.41 \\ \hline
		\multirow{2}{*}{Car in PASCAL} &  \multirow{2}{*}{20} & acc. & \textbf{82.50} & 65.00 & 59.50 & 61.00 \\ \cline{3-7}
		&  & time   & \textbf{0.02} & 0.04 & 1.22 & 0.76 \\ \hline
		\multirow{2}{*}{Motorbike in PASCAL}  &  \multirow{2}{*}{20}  & acc. & \textbf{96.00} & 80.50 & 70.00 & 73.50 \\\cline{3-7}
		&  & time   & \textbf{0.02} & 0.08 & 1.16 & 0.64 \\ \hline\hline
		\multirow{2}{*}{DTU-House}  &  \multirow{2}{*}{500}  & acc. & \textbf{83.62} & 60.64 & - & - \\\cline{3-7}
		&  & time   & \textbf{16} & 711 & - & -    \\ \hline
		\multirow{2}{*}{HerzJesu in DAISY}  &  \multirow{2}{*}{500} & acc.     & \textbf{82.56} & 49.18 & - & - \\\cline{3-7}
		&  & time   & \textbf{11} & 786 & - & -   \\ \hline
		\multirow{2}{*}{Semper in DAISY}  &  \multirow{2}{*}{500} & acc.     & \textbf{92.04} & 54.46 & - & -   \\\cline{3-7}
		&  & time   & \textbf{13} & 619 & - & -   \\ \hline
		\multirow{2}{*}{Stadium in SUIRD}  &  \multirow{2}{*}{500} & acc.    & \textbf{91.68} & 40.46 & - & -   \ \\ \cline{3-7}
		&  & time   & \textbf{10} & 435 & - & -     \\ \hline
	\end{tabular}}
	\caption{Graph matching results. 
    $n =$ number of nodes,
    %Performance with graph matching in real images. 
    acc. $=$ matching accuracy in percent, time $=$ runtime in seconds. }
	\label{tab:GM_real}
\end{table}
\fi

\begin{table}[htb!]
	\centering
	\scalebox{0.9}{
		\begin{tabular}{c||c|c|c|c|c}
			\hline
			Data & Metric & EXPP--II & GNCCP & PATH & LAGSA \\ \hline\hline
			\multirow{2}{*}{DTU-House} & acc. & \textbf{75.70} & 60.10 & 29.10 & 41.60 \\ \cline{2-6} 
			 & time & \textbf{0.51} & 3.57 & 234.55 & 283.32 \\\hline
    \multirow{2}{*}{HerzJesu in DAISY} & acc. & \textbf{78.50} & 58.90 & 22.90 & 35.50 \\\cline{2-6} 
			& time & \textbf{0.50} & 3.71 & 209.89 & 307.61 \\\hline
			\multirow{2}{*}{Fountain in DAISY} & acc. & \textbf{79.50} & 59.00 & 30.90 & 45.10 \\\cline{2-6} 
			 & time & \textbf{0.43} & 2.85 & 325.34 & 278.73 \\\hline
			\multirow{2}{*}{Semper in DAISY} & acc. & \textbf{79.50} & 59.00 & 30.90 & 45.10 \\\cline{2-6} 
			 & time & \textbf{0.50} & 4.42 & 236.76 & 371.07 \\\hline
			\multirow{2}{*}{Brussels in DAISY} & acc. & \textbf{77.80} & 48.40 & 37.80 & 40.40 \\\cline{2-6} 
			 & time & \textbf{0.48} & 3.08 & 207.81 & 302.79 \\\hline
			\multirow{2}{*}{Stadium in SUIRD} & acc. & \textbf{82.10} & 63.10 & 43.00 & 58.20 \\\cline{2-6} 
			& time & \textbf{0.49} & 3.79 & 213.74 & 308.04 \\ \hline
	\end{tabular}}
	\caption{
    Graph matching results in the presence of outliers.
    %Experiment results for graph matching with the presence of outliers. 
    $n= 110$ nodes, $10$ outliers,
    %$n=110$ with $10$ outliers. 
    acc. $=$ matching accuracy in percent, time $=$ runtime in seconds.}
	\label{tab:GM_outlier}
\end{table}

We also want to examine sensitivities with outliers.
We used the same protocol to construct graphs, except that we added $10$ outliers to the graphs.
Table \ref{tab:GM_outlier} shows the average matching accuracies and runtimes.
EXPP--II is seen to show better matching accuracies than the benchmarked algorithms.

\begin{figure*}[htb!]
\centering
\includegraphics[scale=0.45]{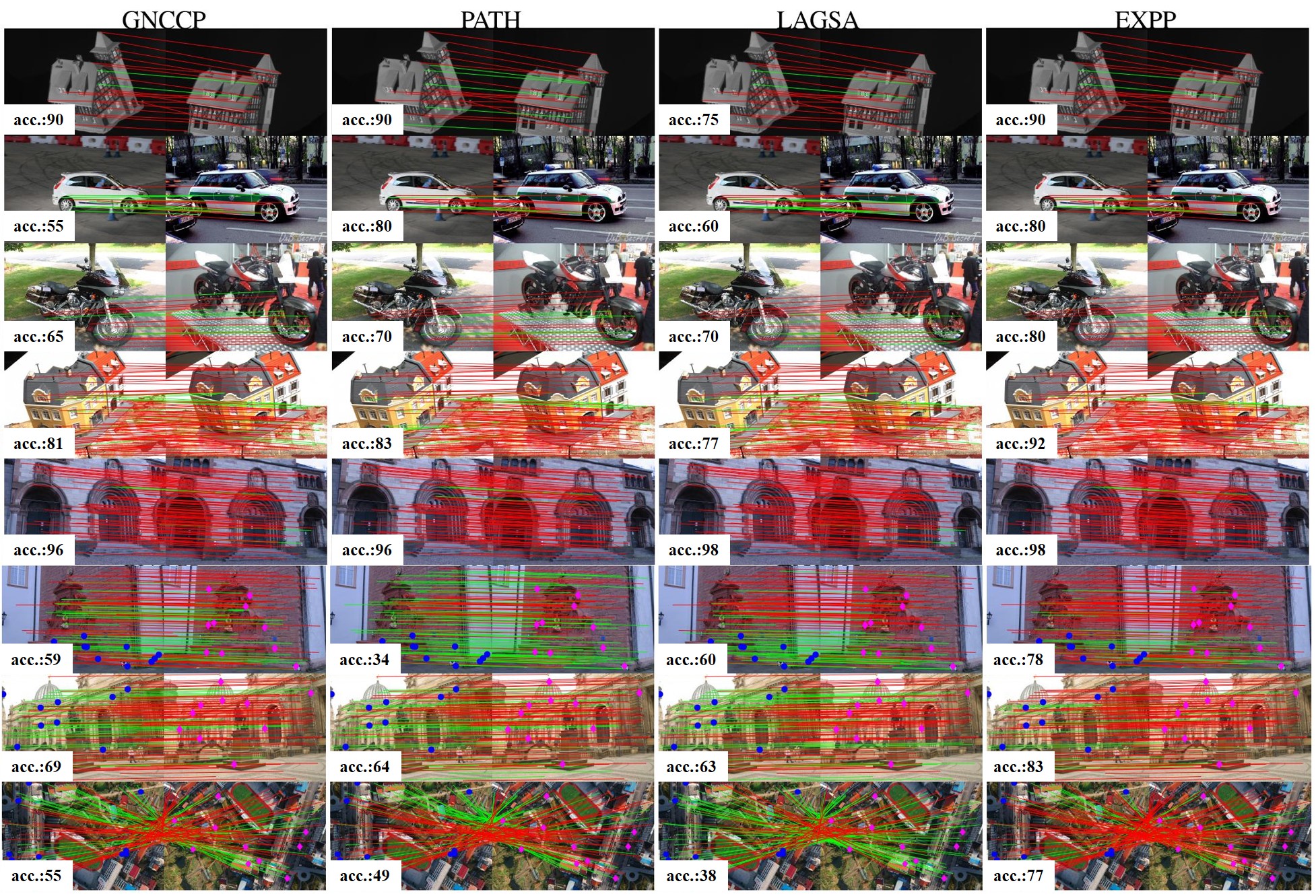}
\caption{{\revMovecolor Illustration of graph matching instances.
First row to third row: CMU-House, Car, Motorbike, with $n=20$;
fourth row to fifth row: DTU-House, HerzJesu, with $n=100$;
sixth row to eighth row: Fountain, Semper, and Stadium, $n=110$, $10$ outliers. 
Red lines: correctly matched nodes; green lines: mismatched nodes.
Blue dots and magenta dots represent outliers in the two figures, respectively.}}
\label{fig:GM_n20100110}
\end{figure*}

\begin{figure*}[htb!]
\centering
\includegraphics[scale=0.45]{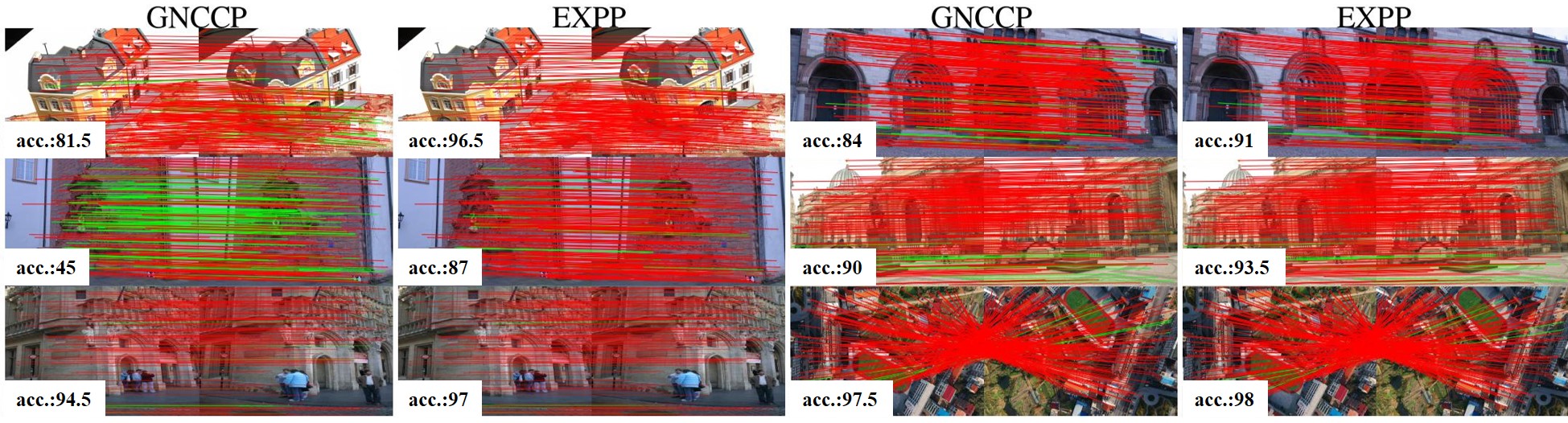}
\caption{{\revMovecolor Illustration of graph matching instances with $n = 200$.
First row: DTU-House, HerzJesu;
second row: Fountain,  Semper;
third row: Brussels,  Stadium. Red lines: correctly matched nodes; green lines: mismatched nodes.}}
\label{fig:GM_n200}
\end{figure*}

{\revMovecolor For visual illustration, we provide matching examples in Fig. \ref{fig:GM_n20100110}--\ref{fig:GM_n200}.
An example for the case of $n=500$ is enlarged and shown in Fig.~\ref{fig:GM_n500_large} for clear visualization.}

%[width=18cm, height=13cm]

%[width=18cm, height=5.1cm]

\ifconfver
\begin{figure}[htb!]
	\centering
	\includegraphics[width=13.06cm, height=8.70cm, angle=90]{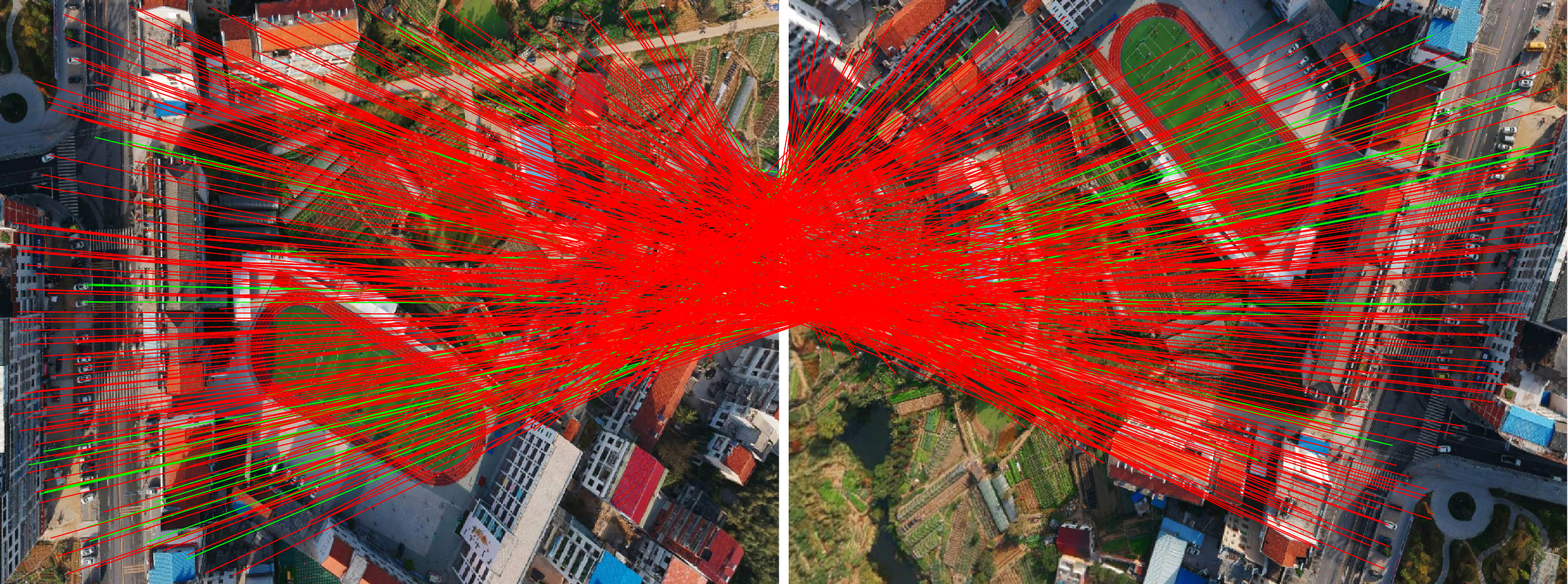}
	\caption{{\revMovecolor A graph matching instance of Stadium with $n = 500$. Red lines: correctly matched nodes; green lines: mismatched nodes.}}
	\label{fig:GM_n500_large}
\end{figure}
\else
\begin{figure}[htb!]
	\centering
	\includegraphics[width=19.2cm, height=12.8cm, angle=90]{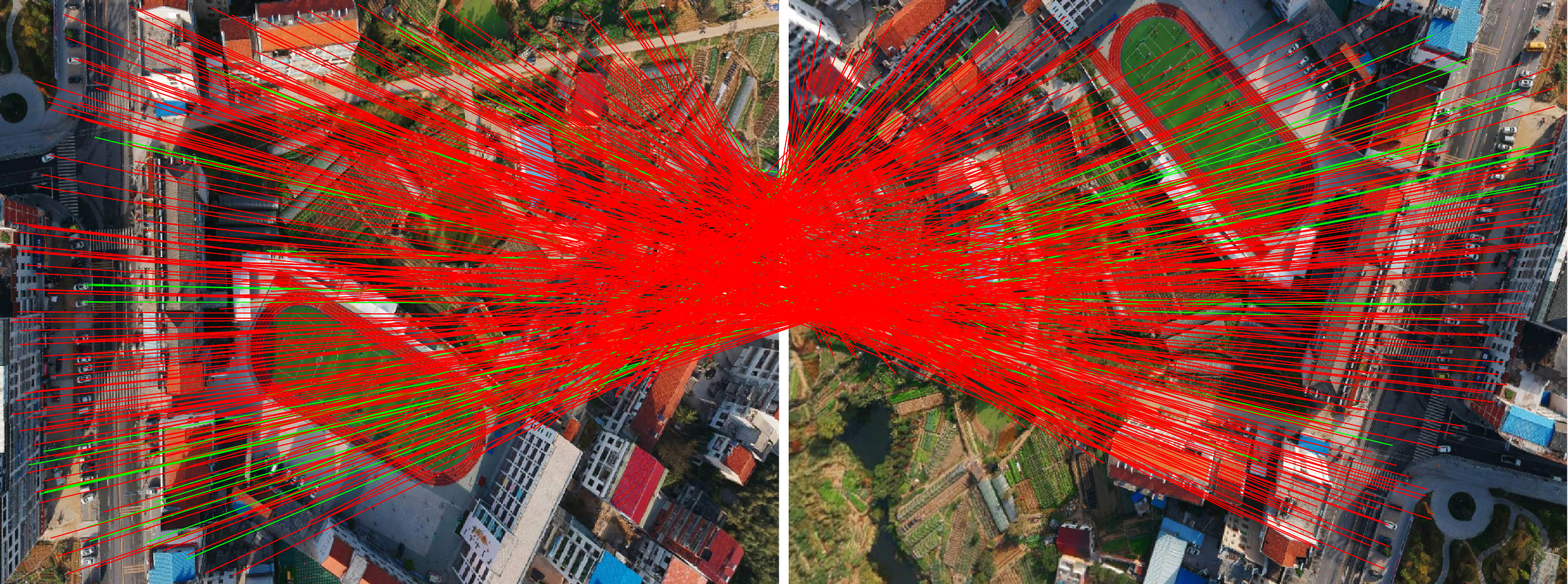}
	\caption{A graph matching instance of Stadium with $n = 500$. Red lines: correctly matched nodes; green lines: mismatched nodes.}
	\label{fig:GM_n500_large}
\end{figure}
\fi

\subsection{Size-Constrained Clustering}
We consider the size-constrained clustering problem described in Section~\ref{sect:clustering_size}.
The benchmarked algorithms are (i) AM: an alternating minimization method \cite{sidiropoulos2015signal};
(ii) SC: $K$-means followed by a post-processing based on mixed-integer linear programming \cite{zhu2010data}; (iii) E-Kmeans: an extension of $K$-means \cite{ganganath2014data}. 
We employ EXPP--II in \eqref{eq:new_expp_sam}.
We revise the objective function (cf. \eqref{eq:prob_scs}) as $f(\bx) +  \gamma \| \bX \|_\fro^2$ with  $\gamma=1.1\sigma_1(\bR)/\kappa_{[r]}$,
%$f(\boldsymbol{X})+\gamma\|\boldsymbol{X}\|_{\text{F}}^2$, with $\gamma=1.1\sigma_1(R)/\kappa_{[r]}$, 
such that the revised objective function is convex  and EXPP--II starts with a convex problem.
%to ensure convexity when $\lambda=0$.
The parameter settings of EXPP--II are $\lambda_0 = 10^{-3}K\nu$,  $\lambda_{k+1} =\lambda_k+0.1K\nu$, $\varepsilon_1 = 10^{-4}$, $\varepsilon_3 = 10^{-2}$, $\bar{L}= 200$, $\bar{\lambda} = K\nu$, where $\nu$ is specified in Theorem \ref{thm:eb2_Unrk} and $K$ is the Lipschitz constant of $f(\boldsymbol{X})$ on $\tilde{\setU}_\bka^{n,r}$. 
%We choose clustering accuracy as the metric.

We evaluate the algorithms on several real-world datasets, some of which are used by the prior work \cite{zhu2010data}.
The results are shown in Table \ref{tab:SCC}.
We see that the clustering accuracies, defined as the number of correctly clustered data points normalized by the total number of data points,
of EXPP--II are generally good compared to the other algorithms.
Note that SC fails to work for the Fashion MNIST dataset, due possibly to the large data size.
EXPP--II cannot compete with E-Kmeans in terms of runtime, but otherwise EXPP--II is much faster than the other algorithms.

\ifconfver
\begin{table*}[htb!]
	\centering
	\scalebox{1}{
\begin{tabular}{c||c|c|c||c|c|c|c|c}
	\hline
	Datasets &   $m$ & $n$ & $\boldsymbol{\kappa}$&  Metric & AM & EXPP--II & SC & E-Kmeans \\ \hline\hline
	\multirow{2}{*}{Fashion MNIST \cite{xiao2017fashion}} & 	\multirow{2}{*}{$784 $} & 	\multirow{2}{*}{$70,000$} & 	\multirow{2}{*}{ $r=10$; $7$K samples/cluster} &acc. & 0.485 & \textbf{0.525} & - & 0.450 \\ \cline{5-9} 
	&&&& time & 7792 & 125 & - & \textbf{51} \\ \hline
	\multirow{2}{*}{Iris \cite{misc_iris_53}} &	\multirow{2}{*}{$4$} & 	\multirow{2}{*}{$150$} & 	\multirow{2}{*}{$(50, 50, 50)$} & acc. & 92.00 & \textbf{93.33} & 81.33 & \textbf{93.33} \\ \cline{5-9} 
	&&&	& time & 0.153 & 0.100 & 0.283 & \textbf{0.006} \\ \hline
	\multirow{2}{*}{Glass Identification \cite{misc_glass_identification_42}} & \multirow{2}{*}{$9 $} & \multirow{2}{*}{$214$} & \multirow{2}{*}{$(70, 76, 17, 13, 9, 29)$} &acc. & 37.38 & 48.60 & \textbf{52.34} & 45.33 \\ \cline{5-9} 
	&&&& time & 0.390 & 0.025 & 0.286 & \textbf{0.007} \\ \hline
	\multirow{2}{*}{Balance Scale \cite{misc_balance_scale_12}} & \multirow{2}{*}{$4$} & \multirow{2}{*}{$625$} & \multirow{2}{*}{$(288, 288, 49)$} & acc. & 60.00 & \textbf{67.52} & 65.60 & 61.44 \\ \cline{5-9} 
	&&&& time & 0.226 & 0.077 & 0.295 & \textbf{0.012} \\ \hline
\end{tabular}}
\caption{Size-constrained clustering results.
%Results of size-constrained clustering on real-world datasets. 
acc. $=$ clustering accuracy in percent, time $=$ runtime in seconds.}
\label{tab:SCC}
\end{table*}
\else
\begin{table*}[htb!]
	\centering
	\scalebox{0.79}{
\begin{tabular}{c||c|c|c||c|c|c|c|c}
	\hline
	Datasets &   $m$ & $n$ & $\boldsymbol{\kappa}$&  Metric & AM & EXPP--II & SC & E-Kmeans \\ \hline\hline
	\multirow{2}{*}{Fashion MNIST \cite{xiao2017fashion}} & 	\multirow{2}{*}{$784 $} & 	\multirow{2}{*}{$70,000$} & 	\multirow{2}{*}{ $r=10$; $7$K samples/cluster} &acc. & 0.485 & \textbf{0.525} & - & 0.450 \\ \cline{5-9} 
	&&&& time & 7792 & 125 & - & \textbf{51} \\ \hline
	\multirow{2}{*}{Iris \cite{misc_iris_53}} &	\multirow{2}{*}{$4$} & 	\multirow{2}{*}{$150$} & 	\multirow{2}{*}{$(50, 50, 50)$} & acc. & 92.00 & \textbf{93.33} & 81.33 & \textbf{93.33} \\ \cline{5-9} 
	&&&	& time & 0.153 & 0.100 & 0.283 & \textbf{0.006} \\ \hline
	\multirow{2}{*}{Glass Identification \cite{misc_glass_identification_42}} & \multirow{2}{*}{$9 $} & \multirow{2}{*}{$214$} & \multirow{2}{*}{$(70, 76, 17, 13, 9, 29)$} &acc. & 37.38 & 48.60 & \textbf{52.34} & 45.33 \\ \cline{5-9} 
	&&&& time & 0.390 & 0.025 & 0.286 & \textbf{0.007} \\ \hline
	\multirow{2}{*}{Balance Scale \cite{misc_balance_scale_12}} & \multirow{2}{*}{$4$} & \multirow{2}{*}{$625$} & \multirow{2}{*}{$(288, 288, 49)$} & acc. & 60.00 & \textbf{67.52} & 65.60 & 61.44 \\ \cline{5-9} 
	&&&& time & 0.226 & 0.077 & 0.295 & \textbf{0.012} \\ \hline
\end{tabular}}
\caption{Size-constrained clustering results.
%Results of size-constrained clustering on real-world datasets. 
acc. $=$ clustering accuracy in percent, time $=$ runtime in seconds.}
\label{tab:SCC}
\end{table*}
\fi

\subsection{Orthogonal {\finalcolor Non-Negative} Matrix Factorization}

We consider the ONMF problem described in Section \ref{sect:ONMF}.
The benchmarked algorithms are 
(i) ONPMF \cite{pompili2014two};
(ii) NSNCP \cite{wang2021clustering}: an alternating minimization algorithm for a penalized ONMF formulation;
(iii) EP4ORTH \cite{jiang2023exact}: a manifold optimization algorithm for the formulation reviewed in \eqref{eq:form_jiang}.
We consider both the EXPP--II formulation in \eqref{eq:new_expp_nsom} and the EXPP formulation in Part I of this paper.
In EXPP we have $h(\boldsymbol{X})=- \|\bX\|_{\text{F}}^2$ and $\mathcal{X}=\setB^{n,r} \cap\mathbb{R}_+^{n \times r}$, and we employ Dykstra's projection algorithm \cite{boyle1986method} to perform projection onto $\setX$.
The parameter settings of EXPP--II (respectively [resp.] EXPP) are $\lambda_0 = 10^{-15}$ (resp. $10^{-5}$), $\lambda_{k+1} = 10\lambda_k$ (resp. $5\lambda_k$), $\varepsilon_1 = 10^{-9}$, $\bar{L}= 50$ (resp. 100), $\bar{\lambda} = K\nu$, where $\nu$ is specified in Theorem \ref{thm:eb2_Snr+} and $K$ is the Lipschitz constant of $f$ on $\tilde{\setB}_+^{n,r}$.
Unlike the previous applications, we found that EXPP and EXPP--II do not work if we start with a $\lambda_0$ such that the problem is convex (the corresponding solution is $\bzero$, which is not a good starting point).
We initialize EXPP and EXPP--II with the NNDSVD method \cite{boutsidis2008svd}, which is commonly adopted by other ONMF methods such as 
%ONPMF \cite{pompili2014two} and EP4ORTH \cite{jiang2023exact}.
ONPMF and EP4ORTH.

We evaluate the algorithms on various real-world datasets: face images
\cite{cai2006orthogonal}, 
hyperspectral images\cite{zhu2014spectral}, and text\cite{cai2010locally}.
Table~\ref{tab:ONMF-datasets} describes the dimensions of these datasets.
The accuracies and runtimes of the various algorithms are shown in Table \ref{tab:ONMF-ACCresults} and Table \ref{tab:ONMF-time}, respectively.
In general, EXPP and EXPP--II perform comparably with the other methods. 
EXPP--II has faster runtimes while EXPP has higher clustering accuracies.

\ifconfver
\begin{table}[htb!]
	\centering
	\scalebox{1}{
		\begin{tabular}{c||c|c|c|c|c|c}
			\hline
			& TDT2 & Reuters & YALE& ORL&Jasper Ridge & Samson \\\hline
			\hline
			$m$ & 34,642 & 5,423 & 1024  & 1,024 & 198 & 156 \\ \hline
			$n$ & 7,809 & 748 & 165  & 150 & 10,000 & 9,025  \\ \hline
			$r$ & 25 & 10 & 15  & 15 & 4 & 3 \\  \hline
	\end{tabular}}
	\caption{Datasets for ONMF.}
	\label{tab:ONMF-datasets}
\end{table}
\else
\begin{table}[htb!]
	\centering
	\scalebox{0.9}{
		\begin{tabular}{c||c|c|c|c|c|c}
			\hline
			& TDT2 & Reuters & YALE& ORL&Jasper Ridge & Samson \\\hline
			\hline
			$m$ & 34,642 & 5,423 & 1024  & 1,024 & 198 & 156 \\ \hline
			$n$ & 7,809 & 748 & 165  & 150 & 10,000 & 9,025  \\ \hline
			$r$ & 25 & 10 & 15  & 15 & 4 & 3 \\  \hline
	\end{tabular}}
	\caption{Datasets for ONMF.}
	\label{tab:ONMF-datasets}
\end{table}
\fi

\begin{table}[!htbp]
	\centering
	\scalebox{0.9}{
		\begin{tabular}{c||c|c|c|c|c|c}
			\hline
			Algorithms & TDT2 & Reuters & YALE & ORL & Jasper Ridge & Samson \\ \hline\hline
			DTPP & 67.20 & 49.20 & \textbf{43.64} & 48.00 & 80.28 & 89.86 \\ \hline
			ONPMF & 69.47 & 50.40 & 35.15 & 42.67 & 79.71 & 94.81 \\ \hline
			NSNCP & 67.44 & 49.20 & 42.42 & \textbf{56.00} & 81.49 & 94.23 \\ \hline
			EP4ORTH & 70.48 & 50.27 & 37.58 & 40.67 & 82.86 & 93.65 \\ \hline
			EXPP--II & 67.38 & 52.14 & 37.58 & 46.67 & 90.27 & 92.72 \\ \hline
			EXPP & \textbf{74.30} & \textbf{64.84} & 37.58 & 42.67 & \textbf{94.46} & \textbf{95.86} \\ \hline
	\end{tabular}}
	\caption{ONMF clustering accuracies in percent.}
	\label{tab:ONMF-ACCresults}
\end{table}

\ifconfver
\begin{table}[!htbp]
	\centering
	\scalebox{0.88}{
		\begin{tabular}{c||c|c|c|c|c|c}
			\hline
			Algorithms & TDT2 & Reuters & YALE & ORL & Jasper Ridge & Samson \\ \hline\hline
			DTPP & 87.46 & 2.19 & \textbf{0.20} & \textbf{0.11} & \textbf{0.63} & \textbf{0.63} \\ \hline
			ONPMF & 1008.67 & 31.94 & 5.02 &  4.38 & 25.10 & 15.90 \\ \hline
			NSNCP & 79.95 & 1.96 & 1.79 & 0.77 & 5.38 & 2.25 \\ \hline
			EP4ORTH & 40.66 & 1.37 & 1.10 &  1.08 & 2.21 & 2.14 \\ \hline
			EXPP--II & \textbf{10.11} & \textbf{0.36} & 0.25 & 0.16 & 2.68 & 2.34 \\ \hline
			EXPP & 63.60 & 2.60 & 0.88 & 0.68 & 8.68 & 6.57 \\ \hline
	\end{tabular}}
	\caption{ONMF runtimes in seconds.}
	\label{tab:ONMF-time}
\end{table}
\else
\begin{table}[!htbp]
	\centering
	\scalebox{0.9}{
		\begin{tabular}{c||c|c|c|c|c|c}
			\hline
			Algorithms & TDT2 & Reuters & YALE & ORL & Jasper Ridge & Samson \\ \hline\hline
			DTPP & 87.46 & 2.19 & \textbf{0.20} & \textbf{0.11} & \textbf{0.63} & \textbf{0.63} \\ \hline
			ONPMF & 1008.67 & 31.94 & 5.02 &  4.38 & 25.10 & 15.90 \\ \hline
			NSNCP & 79.95 & 1.96 & 1.79 & 0.77 & 5.38 & 2.25 \\ \hline
			EP4ORTH & 40.66 & 1.37 & 1.10 &  1.08 & 2.21 & 2.14 \\ \hline
			EXPP--II & \textbf{10.11} & \textbf{0.36} & 0.25 & 0.16 & 2.68 & 2.34 \\ \hline
			EXPP & 63.60 & 2.60 & 0.88 & 0.68 & 8.68 & 6.57 \\ \hline
	\end{tabular}}
	\caption{ONMF runtimes in seconds.}
	\label{tab:ONMF-time}
\end{table}
\fi

% \begin{table}[htb!]
% 	\centering
% 	\scalebox{0.8}{
% 		\begin{tabular}{c|c|c|c|c|c|c}
% 			\hline
% 			Algorithms &TDT2 & Reuters & YALE & ORL&Jasper Ridge & Samson \\
% 			\hline
% 			EXPP-\uppercase\expandafter{\romannumeral2}&2.0e-18&7.1e-18 &2.7e-18 & 3.1e-18&1.7e-17 &2.5e-17\\ \hline
% 			EXPP ($K\nu$)    & 6.9e-8 & 1.9e-6 & 2.1e-6 & 2.3e-6 & 3.8e-7& 5.7e-7 \\ \hline
% 			EXPP ($10K\nu$)  & 4.6e-8 & 1.1e-6 & 6.8e-7 & 9.0e-7 &1.3e-7&1.9e-7 \\ \hline
% 			EXPP ($ 100K\nu $) & 2.7e-8 & 7.3e-7 & 1.2e-7 & 1.7e-7 &5.0e-8 & 5.7e-8 \\ \hline
% 	\end{tabular}}
% 	\caption{Feasibility of EXPP and EXPP--II in ONMF.}
% 	\label{tab:ONMF-feasibility}
% \end{table}

% \begin{table}[!htbp]
% 	\centering
% 	\scalebox{0.9}{
% 		\begin{tabular}{c|c|c|c|c|c|c|c}
% 			\hline
% 			& Metric & TDT2 & Reuters & YALE & ORL & Jasper Ridge & Samson \\ \hline\hline
% 			\multirow{3}{*}{acc.}
% 			& rank & 5    & 2 & 3 & 3 &2 & 5\\  \cline{2-8} 
% 			&acc.  & 67.38 & 52.14 & 37.58 & 46.67 & 90.27 & 92.72 \\  \cline{2-8} 
% 			&gap & -6.92 & -12.70 & -6.06 & -9.33 & -4.19 & -3.14 \\ \hline\hline
% 			\multirow{2}{*}{time}
% 			& rank & 1    & 1 & 2 & 2 & 3 & 3 \\  \cline{2-8}
% 			&time & \textbf{10.11} & \textbf{0.36} & 0.25 & 0.16 & 2.68 & 2.34 \\  \cline{2-8} 
% 			& gap & 0.00 & 0.00 & -0.05 & -0.05 & -2.05 & -1.71 \\ \hline
% 	\end{tabular}}
% 	\caption{\blue Evaluation of the performance of EXPP-\uppercase\expandafter{\romannumeral2}.}
% 	\label{tab:ONMF-EXPPII}
% \end{table}	

\subsection{Sparse and Fair Principal Component Analysis}

We consider sparse and fair PCAs.
Let $\bY \in \Rbb^{m \times n}$ be a data matrix.
Sparse PCA considers 
\[
\min_{\bX \in \setS^{m \times r}} f(\bX) = - \tr(\bX^\top \bR \bX) + \mu \| \bX \|_{\ell_1},
\]
where $\mu > 0$ is given; $\bR = \bY \bY^\top$ is the data correlation matrix.
The aim is to recover sparse principal components.
Let $\bY_t \in \Rbb^{m \times n_t}$, $t=1,\ldots,T$ be data matrices of different groups.
Fair PCA considers 
\[
\min_{\bX \in \setS^{m \times r}} f(\bX) = \max_{t=1,\ldots,T} - \tr(\bX^\top \bR_t \bX),
\]
where $\bR_t = \bY_t \bY_t^\top$ is the data correlation matrix associated with the $t$th group.
The aim of fair PCA is to reduce possible bias to individual groups.
We also consider sparse fair PCA
\[
\min_{\bX \in \setS^{m \times r}} f(\bX) = \max_{t=1,\ldots,T} - \tr(\bX^\top \bR_t \bX) + \mu \| {\revcolor \bX } \|_{\ell_1}.
\]
We use EXPP to handle all the above PCAs.

The benchmarked algorithms for sparse PCA are: 
(i) IMRP: iterative minimization of rectangular
Procrustes \cite{benidis2016orthogonal};
(ii) Gpower-$\ell_1$: generalized power method with $\ell_1$ penalty \cite{journee2010generalized}.
The benchmarked algorithms for fair PCA are
(i) ARPGDA: the alternating Riemannian projected
gradient descent ascent algorithm \cite{xu2022alternating};
(ii) F-FPCA: a subgradient-type algorithm \cite{zalcberg2021fair}.
There is no algorithm to benchmark for sparse fair PCA.

EXPP is implemented by the projected subgradient method since the objective functions are non-smooth.
The parameter settings of EXPP are $\lambda_0 = 0 $, $\lambda_{k+1} = \lambda_k+0.1K $, $\varepsilon_1 = 10^{-3} $, $\varepsilon_3=10^{-6}$, $\bar{L}= 300$, $\bar{\lambda} = K$, where $K$ is the Lipschitz constant of $f$ over $\setB^{n,r}$.
The step size of the projected subgradient method is set as $c/\sqrt{l+1}$ for some $c> 0$.
%Due to the same reason as that in the ONMF problem, we do not start with a $\lambda_0$ such that the objectives are convex.
We initialize EXPP with PCA.
The EXPP for sparse PCA, fair PCA, and sparse fair PCA are called EXPP--S, EXPP--F, and EXPP--SF, respectively.

We use the following performance metrics to evaluate the performance of the various algorithms: explained variance and cardinality, which indicate the trade-off between sparsity and variance; and minimum variance, which measures fairness.
Explained variance is measured as 
\[ {\revcolor 
%\text{\st{$\text{Tr}(\boldsymbol{ X}^\top \bR \boldsymbol{ X})/\text{Tr}(\boldsymbol{\hat X}^\top \bR \boldsymbol{\hat X})$}} 
%\quad 
\tr(\bX^\top \bR \bX )/\tr( \hat{\bX}^\top \bR \hat{\bX}), }
\]
where $\bX$ is the principal component matrix recovered by an algorithm; $\hat{\bX}$ is that of PCA.
Minimum variance is measured as
\[
{\revcolor 
%\text{\st{$\min_{t=1,...,T} \text{Tr}(\boldsymbol{ X}^\top \bR_t \boldsymbol{ X})$}}
%\quad 
\min_{t=1,...,T} \tr( \bX^\top \bR_t \bX ).
}
\]
As a minor step, we project $\bX$ onto $\setS^{n,r}$ before we evaluate the above performance.
This is because some algorithms such as Gpower-$\ell_1$ do not guarantee semi-orthogonality exactly.
%Practically, we project the data to the range space of $\boldsymbol{X}$ since algorithms like Gpower-$\ell_1$ do not guarantee orthogonality exactly. 
The cardinality is measured as the number of elements that are larger than $0.01\max_{i,j}|x_{ij}|$.
We normalize minimum variance and cardinality to $[0,1]$.

We test the algorithms on three real-world datasets: MNIST, Fashion MNIST, and CIFAR10. 
Each dataset consists of $10$ different classes.
The information about the datasets and the corresponding EXPP parameters are provided in Table \ref{tab:str_pca_dataset}.
To create a data matrix, we randomly selected $5$ classes. 
Then, for each class, we randomly selected data points to form one data matrix $\bY_t$.
The data lengths of the different classes are $\{ n_t \} = \{ 5, 10, 500, 1000, 5000 \}$. We consider unbalanced data groups.
We conducted $100$ Monte Carlo runs for each dataset.

The results are shown in Table \ref{tab:str_pca_exp_realdata}.
For sparse PCA, we see that EXPP--S performs better than Gpower-$\ell_1$, while IMRP outperforms EXPP--S.
We note that IMPR adopts a different sparsity penalty. 
For fair PCA, EXPP--F performs comparably with ARPGDA and works better than F-FPCA.
EXPP--SF is seen to be able to strike a balance between sparsity and fairness.
The numerical results demonstrate the versatility of EXPP to handle different formulations.

% PCA results put here to avoid being inserted in the appendix
\ifconfver
\begin{table*}[htb!]
\centering
\scalebox{1}{
\begin{tabular}{c|c|c|c||c|c|c}
\hline
Datasets       & Image Size & Number of Images & Data Type   &EXPP--S  &EXPP--SF  & EXPP--F         \\ \hline
MNIST \cite{lecun2010mnist}        & $28\times28$       & $70000$          & Digits    & $c=0.025$, $\mu=0.3$  &$c=10$, $\mu=0.2$  & $c=0.5$          \\ \hline
Fashion MNIST \cite{xiao2017fashion} & $28\times28$       & $70000$          & Clothing    & $c=0.05$, $\mu=0.25$  &$c=100$, $\mu=0.2$  &$c=20$       \\ \hline
CIFAR10 \cite{krizhevsky2009learning}       & $32\times32$       & $60000$          & real-world objects & $c=0.1$, $\mu=0.25$  &$c=20$, $\mu=0.35$  &$c=20$\\\hline
\end{tabular}}
\caption{Datasets information and EXPP parameter setups. $c$: the constant in stepsize; $\mu$: sparsity penalty parameter.}
    \label{tab:str_pca_dataset}
\end{table*}
\else
\begin{table*}[htb!]
\centering
\scalebox{0.74}{
\begin{tabular}{c|c|c|c||c|c|c}
\hline
Datasets       & Image Size & Number of Images & Data Type   &EXPP--S  &EXPP--SF  & EXPP--F         \\ \hline
MNIST \cite{lecun2010mnist}        & $28\times28$       & $70000$          & Digits    & $c=0.025$, $\mu=0.3$  &$c=10$, $\mu=0.2$  & $c=0.5$          \\ \hline
Fashion MNIST \cite{xiao2017fashion} & $28\times28$       & $70000$          & Clothing    & $c=0.05$, $\mu=0.25$  &$c=100$, $\mu=0.2$  &$c=20$       \\ \hline
CIFAR10 \cite{krizhevsky2009learning}       & $32\times32$       & $60000$          & real-world objects & $c=0.1$, $\mu=0.25$  &$c=20$, $\mu=0.35$  &$c=20$\\\hline
\end{tabular}}
\caption{Datasets information and EXPP parameter setups. $c$: the constant in stepsize; $\mu$: sparsity penalty parameter.}
    \label{tab:str_pca_dataset}
\end{table*}
\fi

\ifconfver
\begin{table*}[htb!]
\centering
\scalebox{0.95}{
\begin{tabular}{c||cccc||cccc||cccc}
\hline
\multirow{2}{*}{Algorithms} & \multicolumn{4}{c||}{MNIST, $r=10$.} & \multicolumn{4}{c||}{Fashion MNIST, $r=15$.} & \multicolumn{4}{c}{CIFAR10, $r=20$.} \\ \cline{2-13} 
 &
  \multicolumn{1}{c|}{Expl. Var.} &
  \multicolumn{1}{c|}{Min. Var.} &
  \multicolumn{1}{c|}{Card.} &
  Time /s&
  \multicolumn{1}{c|}{Expl. Var.} &
  \multicolumn{1}{c|}{Min. Var.} &
  \multicolumn{1}{c|}{Card.} &
  Time /s&
  \multicolumn{1}{c|}{Expl. Var.} &
  \multicolumn{1}{c|}{Min. Var.} &
  \multicolumn{1}{c|}{Card.} &
  Time /s\\ \hline \hline
IMRP &
  \multicolumn{1}{c|}{\textbf{0.7733}} &
  \multicolumn{1}{c|}{0.3149} &
  \multicolumn{1}{c|}{0.1037} &
  0.6130 &
  \multicolumn{1}{c|}{\bf 0.7890} &
  \multicolumn{1}{c|}{0.3093} &
  \multicolumn{1}{c|}{\bf 0.0642} &
  1.6011 &
  \multicolumn{1}{c|}{\bf 0.7126} &
  \multicolumn{1}{c|}{0.5665} &
  \multicolumn{1}{c|}{0.0462} &
  4.0329 \\ \hline
Gpower-$\ell_1$ &
  \multicolumn{1}{c|}{0.6252} &
  \multicolumn{1}{c|}{0.3010} &
  \multicolumn{1}{c|}{0.0998} &
  {\bf 0.4039} &
  \multicolumn{1}{c|}{0.7070} &
  \multicolumn{1}{c|}{0.3141} &
  \multicolumn{1}{c|}{0.1056} &
  1.8938 &
  \multicolumn{1}{c|}{0.4876} &
  \multicolumn{1}{c|}{0.3796} &
  \multicolumn{1}{c|}{\bf 0.0445} &
  4.0616 \\ \hline  
EXPP--S &
  \multicolumn{1}{c|}{0.6439} &
  \multicolumn{1}{c|}{0.2427} &
  \multicolumn{1}{c|}{\bf 0.0834} &
  0.7227 &
  \multicolumn{1}{c|}{0.7579} &
  \multicolumn{1}{c|}{0.2696} &
  \multicolumn{1}{c|}{0.0769} &
  {\bf 0.4915} &
  \multicolumn{1}{c|}{0.6250} &
  \multicolumn{1}{c|}{0.4222} &
  \multicolumn{1}{c|}{0.0517} &
  {\bf 0.6302} \\ \hline \hline
EXPP--SF &
  \multicolumn{1}{c|}{0.4117} &
  \multicolumn{1}{c|}{0.5639} &
  \multicolumn{1}{c|}{0.1505} &
  4.9375 &
  \multicolumn{1}{c|}{0.3336} &
  \multicolumn{1}{c|}{0.5164} &
  \multicolumn{1}{c|}{0.0813} &
  3.9784 &
  \multicolumn{1}{c|}{0.4676} &
  \multicolumn{1}{c|}{0.5221} &
  \multicolumn{1}{c|}{0.1868} &
  3.3913 \\ \hline \hline
EXPP--F &
  \multicolumn{1}{c|}{0.7967} &
  \multicolumn{1}{c|}{\bf 1} &
  \multicolumn{1}{c|}{0.9912} &
  1.9908 &
  \multicolumn{1}{c|}{0.7685} &
  \multicolumn{1}{c|}{\bf 1} &
  \multicolumn{1}{c|}{0.9755} &
  1.9273 &
  \multicolumn{1}{c|}{0.9354} &
  \multicolumn{1}{c|}{0.9996} &
  \multicolumn{1}{c|}{0.9938} &
  \bf 1.0820 \\ \hline
ARPGDA &
  \multicolumn{1}{c|}{0.8143} &
  \multicolumn{1}{c|}{0.9895} &
  \multicolumn{1}{c|}{0.9947} &
  \textbf{0.6747} &
  \multicolumn{1}{c|}{0.8460} &
  \multicolumn{1}{c|}{0.9895} &
  \multicolumn{1}{c|}{0.9948} &
  \bf 0.7944 &
  \multicolumn{1}{c|}{0.9494} &
  \multicolumn{1}{c|}{\bf 1} &
  \multicolumn{1}{c|}{0.9981} &
  1.5184\\ \hline
F-FPCA &
  \multicolumn{1}{c|}{0.9560} &
  \multicolumn{1}{c|}{0.8189} &
  \multicolumn{1}{c|}{1} &
  6.3386 &
  \multicolumn{1}{c|}{0.9490} &
  \multicolumn{1}{c|}{0.8133} &
  \multicolumn{1}{c|}{1} &
  6.9849 &
  \multicolumn{1}{c|}{0.9781} &
  \multicolumn{1}{c|}{0.9484} &
  \multicolumn{1}{c|}{1} &
  4.2289 \\ \hline
\end{tabular}}
\caption{Sparse and/or fair PCA results.}
\label{tab:str_pca_exp_realdata}
\end{table*}
\else
\begin{table*}[htb!]
\centering
\scalebox{0.64}{
\begin{tabular}{c||cccc||cccc||cccc}
\hline
\multirow{2}{*}{Algorithms} & \multicolumn{4}{c||}{MNIST, $r=10$.} & \multicolumn{4}{c||}{Fashion MNIST, $r=15$.} & \multicolumn{4}{c}{CIFAR10, $r=20$.} \\ \cline{2-13} 
 &
  \multicolumn{1}{c|}{Expl. Var.} &
  \multicolumn{1}{c|}{Min. Var.} &
  \multicolumn{1}{c|}{Card.} &
  Time /s&
  \multicolumn{1}{c|}{Expl. Var.} &
  \multicolumn{1}{c|}{Min. Var.} &
  \multicolumn{1}{c|}{Card.} &
  Time /s&
  \multicolumn{1}{c|}{Expl. Var.} &
  \multicolumn{1}{c|}{Min. Var.} &
  \multicolumn{1}{c|}{Card.} &
  Time /s\\ \hline \hline
IMRP &
  \multicolumn{1}{c|}{\textbf{0.7733}} &
  \multicolumn{1}{c|}{0.3149} &
  \multicolumn{1}{c|}{0.1037} &
  0.6130 &
  \multicolumn{1}{c|}{\bf 0.7890} &
  \multicolumn{1}{c|}{0.3093} &
  \multicolumn{1}{c|}{\bf 0.0642} &
  1.6011 &
  \multicolumn{1}{c|}{\bf 0.7126} &
  \multicolumn{1}{c|}{0.5665} &
  \multicolumn{1}{c|}{0.0462} &
  4.0329 \\ \hline
Gpower-$\ell_1$ &
  \multicolumn{1}{c|}{0.6252} &
  \multicolumn{1}{c|}{0.3010} &
  \multicolumn{1}{c|}{0.0998} &
  {\bf 0.4039} &
  \multicolumn{1}{c|}{0.7070} &
  \multicolumn{1}{c|}{0.3141} &
  \multicolumn{1}{c|}{0.1056} &
  1.8938 &
  \multicolumn{1}{c|}{0.4876} &
  \multicolumn{1}{c|}{0.3796} &
  \multicolumn{1}{c|}{\bf 0.0445} &
  4.0616 \\ \hline  
EXPP--S &
  \multicolumn{1}{c|}{0.6439} &
  \multicolumn{1}{c|}{0.2427} &
  \multicolumn{1}{c|}{\bf 0.0834} &
  0.7227 &
  \multicolumn{1}{c|}{0.7579} &
  \multicolumn{1}{c|}{0.2696} &
  \multicolumn{1}{c|}{0.0769} &
  {\bf 0.4915} &
  \multicolumn{1}{c|}{0.6250} &
  \multicolumn{1}{c|}{0.4222} &
  \multicolumn{1}{c|}{0.0517} &
  {\bf 0.6302} \\ \hline \hline
EXPP--SF &
  \multicolumn{1}{c|}{0.4117} &
  \multicolumn{1}{c|}{0.5639} &
  \multicolumn{1}{c|}{0.1505} &
  4.9375 &
  \multicolumn{1}{c|}{0.3336} &
  \multicolumn{1}{c|}{0.5164} &
  \multicolumn{1}{c|}{0.0813} &
  3.9784 &
  \multicolumn{1}{c|}{0.4676} &
  \multicolumn{1}{c|}{0.5221} &
  \multicolumn{1}{c|}{0.1868} &
  3.3913 \\ \hline \hline
EXPP--F &
  \multicolumn{1}{c|}{0.7967} &
  \multicolumn{1}{c|}{\bf 1} &
  \multicolumn{1}{c|}{0.9912} &
  1.9908 &
  \multicolumn{1}{c|}{0.7685} &
  \multicolumn{1}{c|}{\bf 1} &
  \multicolumn{1}{c|}{0.9755} &
  1.9273 &
  \multicolumn{1}{c|}{0.9354} &
  \multicolumn{1}{c|}{0.9996} &
  \multicolumn{1}{c|}{0.9938} &
  \bf 1.0820 \\ \hline
ARPGDA &
  \multicolumn{1}{c|}{0.8143} &
  \multicolumn{1}{c|}{0.9895} &
  \multicolumn{1}{c|}{0.9947} &
  \textbf{0.6747} &
  \multicolumn{1}{c|}{0.8460} &
  \multicolumn{1}{c|}{0.9895} &
  \multicolumn{1}{c|}{0.9948} &
  \bf 0.7944 &
  \multicolumn{1}{c|}{0.9494} &
  \multicolumn{1}{c|}{\bf 1} &
  \multicolumn{1}{c|}{0.9981} &
  1.5184\\ \hline
F-FPCA &
  \multicolumn{1}{c|}{0.9560} &
  \multicolumn{1}{c|}{0.8189} &
  \multicolumn{1}{c|}{1} &
  6.3386 &
  \multicolumn{1}{c|}{0.9490} &
  \multicolumn{1}{c|}{0.8133} &
  \multicolumn{1}{c|}{1} &
  6.9849 &
  \multicolumn{1}{c|}{0.9781} &
  \multicolumn{1}{c|}{0.9484} &
  \multicolumn{1}{c|}{1} &
  4.2289 \\ \hline
\end{tabular}}
\caption{Sparse and/or fair PCA results.}
\label{tab:str_pca_exp_realdata}
\end{table*}
\fi

%-------------------
%--------------------------------------
\section{Conclusion}
\label{sect:p2_con}

As the second part of our study, we developed new EXPP formulations for the cases of PPMs, SAMs, and NSOMs. 
They have lower requirements with constraints and can be efficiently handled from the viewpoint of building algorithms. 
We also demonstrated the utility of EXPP by performing numerical experiments on a variety of applications.

%--------------------------------------
\section{Acknowledgment}
The authors would like to thank Prof. Aritra Konar, Prof. Nikos Sidiropoulos, and Prof. Bo Jiang for sharing their source codes and data sets for the DkS, size-constrained clustering, and
ONMF problems, respectively.

\ifplainver\section*{Appendix} \else \appendix \fi

\ifplainver\renewcommand\thesubsection{\Alph{subsection}} \else \fi

\subsection{Some Basic Results}

We describe some basic results which will be used in this Appendix.
The following lemma encapsulates the proof procedure of Theorems \ref{thm:eb2_Unr}, \ref{thm:eb2_Unrk}, and \ref{thm:eb2_Snr+} in a generic form.
%The following lemma describes an error bound proof approach in an abstract form, and it will be repeatedly used; 
% it captures the essence of the proof procedure for showing the error bounds for the size-constrained assignment matrix set $\setU^{n,r}_\bka$ in the first part of this paper, as well as those for the non-negative semi-orthogonal matrix set {\orange \cite{jiang2023exact,chen2022tight}.}
\begin{Lemma} \label{lem:rep}
	Let $\setX \subseteq \Rbb^n$ be a compact set.
	Let $\setV \subseteq \setX$ a set.
	Suppose that there exists a mapping $\by: \setX \rightarrow \Rbb^n$, a function $h: \setX \rightarrow \Rbb$, and a scalar $\delta > 0$ such that for any $\bx \in \setX$,
	\begin{align}
		\| \bx - \by(\bx) \|_2  & \leq h(\bx), 
		\label{eq:rep_cond1}  \\
		h(\bx) < \delta \quad & \Longrightarrow \quad \by(\bx) \in \setV.
		\label{eq:rep_cond2}
	\end{align}
	Then, we have the inequality
	\beq \label{eq:rep_res}
	\dist(\bx,\setV) \leq \max\left\{ 1, \frac{B}{\delta} \right\}  h(\bx), \quad \forall \bx \in \setX,
	\eeq 
    where $B$ is any constant such that 
	\beq  \label{eq:rep_cond3}
	B \geq \sup \{ \| \bx - \bz \|_2 \mid \bx \in \setX, \bz \in \setV \}.
	\eeq 
\end{Lemma}

{\em Proof of Lemma~\ref{lem:rep}:} \
Let $\bx \in \setX$ be given.
Suppose that $h(\bx) \leq \delta$. 
Then,
$\dist(\bx,\setV) \leq \| \bx - \by(\bx) \|_2 \leq h(\bx)$.
On the other hand, suppose that $h(\bx) > \delta$. 
Then, 
$\dist(\bx,\setV) \leq B \leq B h(\bx)/ \delta$.
Combining the two cases leads to  \eqref{eq:rep_res}.
%the desired result.
\hfill $\blacksquare$
\medskip

It should be noted that the error bound analyses in \cite{jiang2023exact,chen2022tight} essentially introduced the procedure in Lemma~\ref{lem:rep} in a case-specific manner for $\setS_+^{n,r}$.
Our error bound analysis for $\setU^{n,r}_\bka$ in Part I of this paper also used this procedure.
The following results were previously shown and will be used.
\begin{enumerate}[1.]

    \item Unit sphere $\setS^n = \{ \bx \in \Rbb^n \mid \| \bx \|_2 = 1 \}$:
    % The convex hull of $\setS^n$ is $\setB^n = \{ \bx \in \Rbb^n \mid \| \bx \|_2 \leq 1 \}$.
    % Given $\bx \neq \bzero$, the projection of $\bx$ onto $\setS^n$ is $\Pi_\setS^n(\bx) = \bx/\| \bx \|_2$.
    % Given $\bx = \bzero$, 
    % $\Pi_\setS^n(\bx)$ is any unit $\ell_2$-norm vector.
    % For any $\bx \in \setB^n$ we have the error bounds
    Let $\bx \in \Rbb^n$, $\| \bx \|_2 \leq 1$ be given.
    Let $\by = \Pi_{\setS^n}(\bx)$. It holds that 
    \begin{align} \label{eq:err_unit_sphere}
	%\dist(\bx,\setS^n) \leq 1 - \| \bx \|_2 \leq 1 - \| \bx \|_2^2.
    \dist(\bx,\setS^n)  = \| \bx - \by \|_2 \leq 1  - \| \bx \|_2.
	\end{align}
 
	\item Unit vector set $\setU^n = \{ \be_1,\ldots,\be_n \} \subseteq \Rbb^n$:
    Let $\bx \in \Delta^n$ be given. 
    Let $\by = \Pi_{\setU^n}(\bx)$.
    It holds that
    \begin{align} \label{eq:err_U}
	%\dist(\bx,\setU^n) \leq 2( 1 - x_{[1]}) \leq 2( 1 - \| \bx \|_2^2).
    \dist(\bx,\setU^n) = \| \bx - \by \|_2 \leq 2( 1 - \| \bx \|_2^2).
	\end{align}

    \item Selection vector set 
	$\setU_\kappa^n  =  
	\{ \bx \in \{0,1\}^n \mid \bone^\top \bx = \kappa \},$ $\kappa \in \{1,\ldots,n\}$:
    Let 
    $\bx \in \conv(\setU_\kappa^n)$
    %$\bx \in \conv(\setU_\kappa^n) = \{ \bx \in [0,1]^n \mid \bone^\top \bx = \kappa \}$ 
    be given.
    Let $\by = \Pi_{\setU_\kappa^n}(\bx)$.
    It holds that 
	% The convex hull of $\setU_\kappa^n$ is 
 %    %\[
 %    $\conv(\setU_\kappa^n) = \{ \bx \in [0,1]^n \mid \bone^\top \bx = \kappa \}.$
 %    %\]
 %    For any $\bx \in \conv(\setU_\kappa^n)$ we have the error bounds
    \begin{align} \label{eq:err_Uk}
	% \dist(\bx,\setU^n_\kappa) \leq 2( \kappa - s_\kappa(\bx)) \leq 2( \kappa - \| \bx \|_2^2).
    \dist(\bx,\setU^n_\kappa) = \| \bx - \by \|_2 \leq 2( \kappa - \| \bx \|_2^2).
	\end{align}

    \item Scaled unit vector set
    $\setW^n = \{ \bx \in \Rbb^n \mid \bx = \alpha \bu, \alpha \in \Rbb, \bu \in \setU^n \}$:
    Let $\bx \in \Rbb^n$ be given.
    Let $\by = \Pi_{\setW^n}(\bx) = x_l \be_l$, where $l$ is such that $|x_l|= \| \bx \|_\infty$.
    It holds that 
    \begin{align} \label{eq:err_Wn}
    \dist(\bx,\setW^n) = \| \bx - \by \|_2 \leq 
    % \| \bx - \by \|_1 
     \| \bx \|_1 - \| \bx \|_\infty.
	\end{align}

 \end{enumerate}

\subsection{Proof of Theorem~\ref{thm:eb2_Unr}}

\label{app:thm:eb2_Unr}
% \medskip
% {\em Proof of Theorem~\ref{thm:eb2_Unr}:} \

Let $\bX \in \Rbb^{n \times r}$, with $\bx_j \in \Delta^n$ for all $j$, be given.
Let $\bY \in \Rbb^{n \times r}$ be given by 
\beq \label{eq:y_eb2_Unr}
\by_j = \Pi_{\setU^n}(\bx_j), \quad j=1,\ldots,r.
\eeq 
Let 
\beq \label{eq:h_eb2_Unr}
h(\bX) = 2 \| \bX^\top \bX - \bI \|_{\ell_1}.
\eeq 
We want to apply Lemma~\ref{lem:rep}. 
We first show the first condition \eqref{eq:rep_cond1} of Lemma~\ref{lem:rep}.
Using the error bound result \eqref{eq:err_U} for $\setU^n$, 
the error $\| \bX - \bY \|_\fro$ is bounded as
\[
\| \bX - \bY \|_{\rm F} \leq \sum_{j=1}^r \| \bx_j - \by_j \|_2 \leq 2 (r - \| \bX \|_{\rm F}^2).
\]
From the proof of Lemma~\ref{lem:q_identity} (see Section \ref{sect:eb_ppm}), it can be seen that
$\| \bX^\top \bX - \bI \|_{\ell_1} \geq r - \| \bX \|_\fro^2$.
It follows that the choice in \eqref{eq:y_eb2_Unr}--\eqref{eq:h_eb2_Unr} satisfies the first condition \eqref{eq:rep_cond1} of Lemma~\ref{lem:rep}.

Second we show that the second condition \eqref{eq:rep_cond2} of Lemma~\ref{lem:rep} is satisfied for some $\delta$.
Consider the following lemma.
\begin{Lemma} \label{lem:eb2_Unr}
	Let $\bY \in \{ 0,1 \}^{n \times r}$ be a matrix with $\by_j \in \{ \be_1,\ldots,\be_n \}$ for all $j$ and with $n \geq r$.
	The singular values of $\bY$ satisfy $\sigma_i(\bY)^2 \in \{0,1,\ldots,r\}$ for all $i$ and $\sum_{i=1}^r \sigma_i(\bY)^2 = r$.
\end{Lemma}

{\em Proof of Lemma~\ref{lem:eb2_Unr}:} \
Represent each $\by_j$ by $\by_j = \be_{l_j}$ for some $l_j \in \{ 1,\ldots,n \}$.
We have 
\[
\bY \bY^\top = \textstyle \sum_{j=1}^r \be_{l_j} \be_{l_j}^\top = \Diag(\bd),
\]
where $d_i \in \{0,1,\ldots,r\}$ counts the number of times $\be_i$ appears in $\by_1,\ldots,\by_r$.
As a basic SVD result,
the above equation implies that 
%$\lambda_i(\bY\bY^\top) = d_{[i]}$, and that $\sum_{i=1}^r \lambda_i(\bY\bY^\top) = r$.
$\sigma_i(\bY)^2 = d_{[i]}$ for all $i$.
%Applying $\lambda_i(\bY\bY^\top) = \sigma_i(\bY)^2$ gives the desired result.
Since $\sum_{i=1}^r d_i = r$, it follows that $\sum_{i=1}^r \sigma_i(\bY)^2 = r$.
\hfill $\blacksquare$
\medskip

Lemma~\ref{lem:eb2_Unr} implies that, if $\sigma_i(\bY) < \sqrt{2}$ for all $i$, then $\sigma_1(\bY) = \cdots = \sigma_r(\bY) = 1$ and consequently $\bY$ lies in $\setU^{n,r}$.
Suppose that $h(\bX) < \delta$ for some $\delta > 0$.
By the Weyl inequality,
\ifconfver
\begin{align}\label{eq:eb2_Unr_proof1}
	{\revcolor\sigma_i(\bY)} & {\revcolor\leq \sigma_i(\bX) + \sigma_1(\bY-\bX) 
	%\nonumber  \\
	%&
	\leq \sigma_i(\bX) + \| \bY-\bX \|_{\rm F} \nonumber} \\
	& {\revcolor\leq \sigma_i(\bX) + h(\bX)} \nonumber  \\
	& {\revcolor< \sigma_i(\bX)  + \delta.}
\end{align}
\else
\begin{equation}\label{eq:eb2_Unr_proof1}
    {\revcolor \sigma_i(\bY)  \leq \sigma_i(\bX) + \sigma_1(\bY-\bX) 
	%\nonumber  \\
	%&
	\leq \sigma_i(\bX) + \| \bY-\bX \|_{\rm F}  
	 \leq \sigma_i(\bX) + h(\bX)   
	 < \sigma_i(\bX)  + \delta.} 
\end{equation}
\fi
Also, it is seen from \eqref{eq:h_eb2_Unr} that
\ifconfver
\begin{align*}
	{\revcolor\tfrac{1}{2} h(\bX)} & {\revcolor\geq \| \bX^\top \bX - \bI \|_{\rm F} = \| \bsig(\bX)^2 - \bone \|_2}  \\
	& {\revcolor\geq | \sigma_i(\bX)^2 - 1 |,}
\end{align*}
\else
$$
{\revcolor	\tfrac{1}{2} h(\bX)  \geq \| \bX^\top \bX - \bI \|_{\rm F} = \| \bsig(\bX)^2 - \bone \|_2  \\
	 \geq | \sigma_i(\bX)^2 - 1 |,}
$$
\fi
for any $i$.
The above equation implies that $\sigma_i(\bX) < \sqrt{1 + \delta/2}$.
Applying this to \eqref{eq:eb2_Unr_proof1} yields
\[
\sigma_i(\bY) < \sqrt{1+\delta/2} + \delta \leq 1 + \delta/2 + \delta.
\]
By choosing $\delta = 2(\sqrt{2} - 1)/3$ such that $\sigma_i(\bY) < \sqrt{2}$ for all $i$,
the matrix $\bY$ must lie in $\setU^{n,r}$.
Hence, the second condition \eqref{eq:rep_cond2} of Lemma~\ref{lem:rep} is satisfied for the choice of $h$ and $\by$ in \eqref{eq:y_eb2_Unr}--\eqref{eq:h_eb2_Unr} and for $\delta = 2(\sqrt{2} - 1)/3$.

With the two conditions of Lemma~\ref{lem:rep} satisfied, we are ready to apply Lemma~\ref{lem:rep}.
It can be verified that $B= \sqrt{2r}$ satisfies the requirement \eqref{eq:rep_cond3} in Lemma~\ref{lem:rep} (see (36) in Part I of this paper).
%which, as we have previously shown (cf. Section~\ref{sect:err_bnd_illus}), satisfies the requirement \eqref{eq:rep_cond3}.
%It was previously shown that $\sup \{ \| \bX - \bZ \|_{\rm F} \mid \bX \in \setX, \bZ \in \setU^{n,r} \} \leq \sqrt{2r}$.
Applying the above $h$, $\delta$ and $B$ to \eqref{eq:rep_res} in Lemma~\ref{lem:rep}, we get
\begin{align*}
	\dist(\bX,\setU^{n,r}) & \leq \frac{ 2 \sqrt{2}  }{2(\sqrt{2} - 1)/3} \sqrt{r} \| \bX^\top \bX - \bI \|_{\ell_1}.
\end{align*}
The above inequality leads to the final result \eqref{eq:eb2_Unr_a}; note $(2 \sqrt{2})/[ 2(\sqrt{2} - 1)/3 ] = 10.2426$.
The accompanied result \eqref{eq:eb2_Unr_b} is obtained by applying Lemma~\ref{lem:q_identity} to \eqref{eq:eb2_Unr_a}.

%-------------------
\subsection{Proof of Theorem~\ref{thm:eb2_Unrk}}
\label{app:thm:eb2_Unrk}

Assume that $\ka_1 \geq \cdots \geq \ka_r$ without loss of generality.
Let $\bX \in \Rbb^{n \times r}$, with $\bx_j \in \conv(\setU_{\ka_j}^n)$ for all $j$, be given.
Let $\bY \in \Rbb^{n \times r}$ be given by 
\beq \label{eq:y_eb2_Unrk}
\by_j = \Pi_{\setU^n_{\ka_j}}(\bx_j), \quad j=1,\ldots,r.
\eeq 
Let
\beq \label{eq:h_eb2_Unrk}
h(\bX) = 2 \| \bX^\top \bX - \Diag(\bka) \|_{\ell_1}.
\eeq 
Once again we apply Lemma~\ref{lem:rep}.
Following the same proof as Theorem~\ref{thm:eb2_Unr}, it can be shown that
\[
\| \bX - \bY \|_\fro \leq 2 (\bone^\top \bka - \| \bX \|_\fro^2 ) \leq h(\bX),
\]
where the first inequality is due to the error bound result \eqref{eq:err_Uk} for $\setU^n_\ka$;
the second inequality can be seen from the proof of Lemma~\ref{lem:q_identity}.
The first condition \eqref{eq:rep_cond1} of Lemma~\ref{lem:rep} is thereby satisfied.
The second condition \eqref{eq:rep_cond2} of Lemma~\ref{lem:rep} is more challenging to show.
Consider the following lemma.
\begin{Lemma} \label{lem:eb2_Unrk}
	Let $\bY \in \{ 0,1 \}^{n \times r}$ be a matrix with $\by_j \in \setU_{\ka_j}^n$ for all $j$ and with $n \geq r$.
	If $\sum_{i=1}^r \sigma_i(\bY)^4 < \bone^\top (\bka^2) + 2$, then it must be true that $\by_i^\top \by_j = 0$ for all $i \neq j$.
	Consequently, the matrix $\bY$ lies in $\setU_{\bka}^{n,r}$.
\end{Lemma}

{\em Proof of Lemma~\ref{lem:eb2_Unrk}:} \
Consider the matrix product $\bY^\top \bY$. 
Let $i, j \in \{ 1,\ldots,r \}$, $i \neq j$, be any two distinct indices.
On the one hand,
\ifconfver
\begin{align*}
{\revcolor\| \bY^\top \bY \|_{\rm F}^2} & {\revcolor\textstyle \geq \sum_{l=1}^r \| \by_l \|_2^4 + 2 ( \by_i^\top \by_j )^2} \\
& {\revcolor= \bone^\top ( \bka^2 ) + 2 ( \by_i^\top \by_j )^2.}
\end{align*}
\else
$$
{\revcolor \| \bY^\top \bY \|_{\rm F}^2  \textstyle \geq \sum_{l=1}^r \| \by_l \|_2^4 + 2 ( \by_i^\top \by_j )^2 
 = \bone^\top ( \bka^2 ) + 2 ( \by_i^\top \by_j )^2.}
$$
\fi
On the other hand, 
\[
\| \bY^\top \bY \|_{\rm F}^2 = \| \bm \bsig(\bY)^2 \|_2^2= 
\textstyle \sum_{i=1}^r \sigma_i(\bY)^4.
\]
The above derivations imply that
\[
( \by_i^\top \by_j )^2 \leq \tfrac{1}{2} \textstyle [ \sum_{i=1}^r \sigma_i(\bY)^4 - \bone^\top ( \bka^2 ) ].
\]
Since $\bY \in \{ 0,1 \}^{n \times r}$, we have $\by_i^\top \by_j \in \{ 0, 1,\ldots,n \}$.
If $\sum_{i=1}^r \sigma_i(\bY)^4 < \bone^\top (\bka^2) + 2$, then we are left with $\by_i^\top \by_j = 0$.
The proof is complete.
\hfill $\blacksquare$
\medskip

Suppose that $h(\bX) < \delta$ for some $\delta > 0$.
By the same proof as before (cf. \eqref{eq:eb2_Unr_proof1}), 
\begin{align}
	\sigma_i(\bY) 
	& < \sigma_i(\bX)  + \delta. \label{eq:eb2_Unrk_proof1}
\end{align}
Also, from \eqref{eq:h_eb2_Unrk},
\ifconfver
\begin{align*}
	{\revcolor\tfrac{1}{2} h(\bX)} & {\revcolor\geq \| \bX^\top \bX - \Diag(\bka) \|_{\fro} \geq \| \bsig(\bX)^2 - \bka \|_2}  \\
	& {\revcolor\geq | \sigma_i(\bX)^2 - \ka_i |},
\end{align*}
\else
$$
{\revcolor	\tfrac{1}{2} h(\bX)  \geq \| \bX^\top \bX - \Diag(\bka) \|_{\fro} \geq \| \bsig(\bX)^2 - \bka \|_2 
	 \geq | \sigma_i(\bX)^2 - \ka_i |,}
$$
\fi
for any $i$. 
Here, the second inequality is due to the von Neumann trace inequality $\| \bA - \bB \|_\fro \geq \| \bsig(\bA) - \bsig(\bB) \|_2$.
The above inequality implies $\sigma_i(\bX) < \sqrt{\ka_i + \delta/2}$.
Putting it into \eqref{eq:eb2_Unrk_proof1} leads to
\begin{align}
	\sigma_i(\bY)^2 & < \ka_i + \delta/2 + 2 \delta \sqrt{\ka_i + \delta/2} + \delta^2 \nonumber \\
	& \leq \ka_i + \delta/2 + 2 \delta [ \sqrt{\ka_i} + \delta/(2\sqrt{\ka_i}) ] + \delta^2 \nonumber \\
	& = \ka_i + ( 1/2 + 2 \sqrt{\ka_i} ) \delta + ( 1 + 1/\sqrt{\ka_i} ) \delta^2, \label{eq:eb2_Unrk_proof2}
\end{align}
where the second equation is due to the inequality $\sqrt{a+b} \leq \sqrt{a} + b/\sqrt{a}$ for $a > 0$ and $b \geq 0$.
Suppose that 
\beq \label{eq:eb2_Unrk_proof3}
( 1 + 1/\sqrt{\ka_i} )^\half \delta \leq 1, \quad \forall i.
\eeq 
Eq.~\eqref{eq:eb2_Unrk_proof2} can be further bounded as
\begin{align} \label{eq:eb2_Unrk_proof4}
	\sigma_i(\bY)^2 & < 
	\ka_i + \underbrace{[ 1/2 + 2 \sqrt{\ka_i} + ( 1 + 1/\sqrt{\ka_i} )^\half ]}_{:= \alpha_i} \delta, 
\end{align}
Taking square on \eqref{eq:eb2_Unrk_proof4} gives
\begin{align} \label{eq:eb2_Unrk_proof5}
	\sigma_i(\bY)^4 & < 
	\ka_i^2 + 2 \ka_i \alpha_i \delta + \alpha_i^2 \delta^2.
\end{align}
Suppose that 
\beq \label{eq:eb2_Unrk_proof6}
\alpha_i \delta \leq 1, \quad \forall i.
\eeq 
Then, we can further bound \eqref{eq:eb2_Unrk_proof5} as
\begin{align} \label{eq:eb2_Unrk_proof7}
	\sigma_i(\bY)^4 & < 
	\ka_i^2 + \underbrace{(2 \ka_i \alpha_i + \alpha_i )}_{:= \beta_i} \delta, 
\end{align}
and consequently,
\[
\textstyle
\sum_{i=1}^r \sigma_i(\bY)^4  < \bone^\top (\bka^2) + (\bone^\top \bbeta) \delta.
\]
Let us choose 
$$\delta = 2/(\bone^\top \bbeta).$$ 
It can be verified that this choice of $\delta$ satisfies the requirements \eqref{eq:eb2_Unrk_proof3} and \eqref{eq:eb2_Unrk_proof6}.
By invoking Lemma~\ref{lem:eb2_Unrk}, we see that $\bY$ lies in $\setU_{\bka}^{n,r}$.
Hence, the second condition \eqref{eq:rep_cond2} of Lemma~\ref{lem:rep} is satisfied.

Finally, we assemble the components together to obtain the error bound.
Let $B = \sqrt{2 \bone^\top \bka }$, which can be verified to satisfy \eqref{eq:rep_cond3}.
%It was previously shown that $\sup \{ \| \bX - \bZ \|_{\rm F} \mid \bX \in \setX, \bZ \in \setU^{n,r}_{\bka} \} \leq \sqrt{2 \bone^\top \bka } \leq 1.5 \sqrt{\bone^\top \bka } = B$.
We express $\beta_i$ as
\ifconfver
\begin{align*}
{\revcolor\beta_i} & {\revcolor= (1+ 2\ka_i) [ 1/2 + 2 \sqrt{\ka_i} + \underbrace{( 1 + 1/\sqrt{\ka_i} )^\half}_{\leq \sqrt{2} \leq 1.5} ]} \\
& {\revcolor\leq  2 (1+ 2\ka_i)( 1 + \sqrt{\ka_i} ).}
\end{align*}
\else
$$
{\revcolor \beta_i  = (1+ 2\ka_i) [ 1/2 + 2 \sqrt{\ka_i} + \underbrace{( 1 + 1/\sqrt{\ka_i} )^\half}_{\leq \sqrt{2} \leq 1.5} ]
 \leq  2 (1+ 2\ka_i)( 1 + \sqrt{\ka_i} ).}
$$
\fi
We have
\[
\frac{B}{\delta} \leq \sqrt{2 \bone^\top \bka } ( \underbrace{\textstyle \sum_{j=1}^r (1+ 2\ka_j)( 1 + \sqrt{\ka_j} )}_{:=\gamma}).
\]
Note that the right-hand side of the above equation is greater than $1$.
Applying the above $h$, $\delta$, and $B$ to \eqref{eq:rep_res} in Lemma~\ref{lem:rep} leads to
\begin{align*}
	\dist(\bX,\setU^{n,r}) & \leq 2 \sqrt{2 \bone^\top \bka} \gamma  \| \bX^\top \bX - \Diag(\bka) \|_{\ell_1},
\end{align*}
%where $\gamma = 3 \sum_{i=1}^r (1+ 2\ka_i)( 1 + \sqrt{\ka_i} )$.
and consequently the final result \eqref{eq:eb2_Unrk_a}.
In addition, applying Lemma~\ref{lem:q_identity} to \eqref{eq:eb2_Unrk_a} gives \eqref{eq:eb2_Unrk_b}.

\subsection{Proof of Theorem~\ref{thm:eb2_Snr+}}
\label{app:thm:eb2_Snr+}

Let $\bX \in \Rbb^{n \times r}_+$, with $\| \bx_j \|_2 \leq 1$ for all $j$, be given.
Let $\bW \in \Rbb^{n \times r}$ be given by $\bar{\bw}_i = x_{i,l_i} \be_{l_i}$ for all $i$,
where $l_i$ is such that $x_{i,l_i} = \max\{ x_{i,1},\ldots,x_{i,r} \}$.
Let $\bY \in \Rbb^{n \times r}$ be given by 
\[
\by_j = \left\{
	\begin{array}{ll}
		\bw_j/\| \bw_j \|_2, & \bw_j \neq \bzero, \\
		\bu, & \bw_j = \bzero,
	\end{array} 
\right. 
\]
for all $j$, where $\bu$ denotes any non-negative unit $\ell_2$ norm vector.
It is worth noting that
\[
\bar{\bw}_i = \Pi_{\setW^r}(\bar{\bx}_i), \quad
\by_j = \Pi_{\setS^n}(\bw_j),
\]
for all $i,j$.
Let 
\beq \label{eq:eb2_S+_mark1}
h(\bX) = \sqrt{6} \left[ \sum_{j=1}^r c_1(\bx_j)^2 + \sum_{i=1}^n \rho_1(\bar{\bx}_i)^2 \right]^\half,
\eeq 
where $c_1(\bx) = 1 - \| \bx \|_2$, $\rho_1(\bx) = \| \bx \|_1 - \| \bx \|_\infty$.
As before we apply Lemma~\ref{lem:rep}.
We begin by showing the first condition \eqref{eq:rep_cond1} of Lemma~\ref{lem:rep}. 
Using the error bounds \eqref{eq:err_Wn} and \eqref{eq:err_unit_sphere} for $\setW^r$ and $\setS^n$, respectively, it holds that
\begin{align}
    \| \bar{\bx}_i - \bar{\bw}_i \|_2 & \leq \rho_1(\bar{\bx}_i), 
    \label{eq:eb2_S+_tt1}  \\
    \| \bw_j - \by_j \|_2 & \leq 1 - \| \bw_j \|_2 \nonumber \\
    & \leq 1 - \| \bx_j \|_2 + \| \bw_j - \bx_j \|_2 
    \label{eq:eb2_S+_tt2} \\
    & \leq \sqrt{2} \left[ c_1(\bx_j)^2 + \| \bw_j - \bx_j \|_2^2  \right]^\half,
    \label{eq:eb2_S+_tt3}
\end{align}
where \eqref{eq:eb2_S+_tt2} is due to the triangle inequality; \eqref{eq:eb2_S+_tt3} is due to $(|a|+|b|)^2 \leq 2 ( |a|^2 + |b|^2 )$.
This gives rise to 
\begin{align}
    \| \bX - \bY \|_\fro & \leq  \| \bX - \bW \|_\fro + \| \bW - \bY \|_\fro \nonumber \\
    & \leq \sqrt{2} \left( \| \bX - \bW \|_\fro^2 +  \| \bW - \bY \|_\fro^2 \right)^\half \nonumber \\
    & \leq \sqrt{2} \left[ 3 \| \bX - \bW \|_\fro^2 +  2 {\textstyle \sum_{j=1}^r c_1(\bx_j)^2 } \right]^\half \nonumber \\
    & \leq \sqrt{6} \left[ {\textstyle \sum_{i=1}^n \rho_1(\bar{\bx}_i)^2 } +  {\textstyle \sum_{j=1}^r c_1(\bx_j)^2 } \right]^\half 
    \nonumber
    \\
    & = h(\bX), \label{eq:eb2_S+_tt4}
\end{align}
where the third inequality is due to \eqref{eq:eb2_S+_tt3};
the fourth inequality is due to \eqref{eq:eb2_S+_tt1}.
The first condition \eqref{eq:rep_cond1} of Lemma~\ref{lem:rep} is obtained.

The proof of the second condition \eqref{eq:rep_cond2} of Lemma~\ref{lem:rep} is as follows.
From the way $\bY$ is constructed, we notice that $\bY$ is a non-negative semi-orthogonal matrix if $\| \bw_j \|_2 > 0$ for all $j$.
Suppose that $h(\bX) < \delta$ for some $\delta > 0$.
As a reverse version of \eqref{eq:eb2_S+_tt4}, we have, for any $j$,
\begin{align*}
    \delta & > \sqrt{6} \left[ \| \bX - \bW \|_\fro^2 +  \sum_{j=1}^r c_1(\bx_j)^2 \right]^\half \\
    & \geq \sqrt{6} \left[ \| \bx_j - \bw_j \|_2^2 +  c_1(\bx_j)^2 \right]^\half \\
    & \geq \sqrt{3} \left[ \| \bx_j - \bw_j \|_2 +  c_1(\bx_j) \right] \\
    & = \sqrt{3} \left( \| \bx_j - \bw_j \|_2 +  1 - \| \bx_j \|_2 \right) \\
    & \geq  \sqrt{3} \left(  1 - \| \bw_j \|_2 \right).
\end{align*}
Here, the third inequality is due to $(|a|+|b|)^2 \leq 2 ( |a|^2 + |b|^2 )$ and $c_1(\bx_j) \geq 0$ for any $\| \bx_j \|_2 \leq 1$; the last inequality is due to the triangle inequality.
The above equation implies that, if $\delta = \sqrt{3}$, then $\| \bw_j \|_2 > 0$.
The second condition \eqref{eq:rep_cond2} of Lemma~\ref{lem:rep} is shown.

Finally, we put together the components.
We choose $B= \sqrt{2r}$, which can be verified to satisfy \eqref{eq:rep_cond3}.
The inequality \eqref{eq:rep_res} associated with the above $\delta$, $h$, and $B$ is 
\beq  \label{eq:eb2_S+_mark3}
\dist(\bX,\setS_+^{n,r}) \leq \nu \underbrace{\left[ \sum_{j=1}^r c_1(\bx_j)^2 + \sum_{i=1}^n \rho_1(\bar{\bx}_i)^2 \right]^\half}_{=\psi_2(\bX)},
\nonumber 
\eeq 
where $\nu = \max\{ \sqrt{6}, 2 \sqrt{r} \}$.
Furthermore, as a basic norm result, we have $\psi_2(\bX) \leq \sum_{j=1}^r c_1(\bx_j) + \sum_{i=1}^n \rho_1(\bar{\bx}_i) = \psi_1(\bX)$.
The proof of  Theorem~\ref{thm:eb2_Snr+} is complete.

\subsection{Relationship of the Proof of Theorem~\ref{thm:eb2_Snr+} with Prior Work}
\label{app:diff_onmf_er}

A key step of the proof of Theorem~\ref{thm:eb2_Snr+} 
in Appendix \ref{app:thm:eb2_Snr+}
is the construction of $\bW$ and $\bY$.
This step is identical to that in the prior studies \cite{jiang2023exact,chen2022tight}.
%the prior study \cite{chen2022tight} and its predecessor \cite{jiang2023exact}.
The important difference lies in the bound in \eqref{eq:eb2_S+_tt1}.
We use the error bound for scaled unit vectors, given in Lemma~\ref{lem:setW}, to derive the bound in \eqref{eq:eb2_S+_tt1}.
The prior studies used another bound:
For any $\bX \in \Rbb^{n \times r}_+$, it holds that 
\beq \label{eq:eb2_S+_relate}
\| \bX - \bW \|_\fro^2 \leq \sum_{j=1}^r \sum_{l=1, l \neq j}^r \bx_j^\top \bx_l = \| \bX \bone \|_2^2 - \| \bX \|_\fro^2;
\eeq 
see \cite[Lemma 3.1]{jiang2023exact} and \cite[Lemma 7]{chen2022tight}.
An oversimplified way to understand the prior studies is as follows: replace our bound \eqref{eq:eb2_S+_tt1} with \eqref{eq:eb2_S+_relate}, and then proceed with the error bound proof in Appendix \ref{app:thm:eb2_Snr+}.
There are differences with the fine details and the reader is referred to the prior studies \cite{jiang2023exact,chen2022tight} for a thorough comparison.
%Actually there are many differences but they are mostly with the fine details of the derivations.
Our bound \eqref{eq:eb2_S+_tt1} may be seen {\revcolor as} an improvement over \eqref{eq:eb2_S+_relate}.
From \eqref{eq:eb2_S+_tt1} and the proof of Lemma
\ref{cor:eb2_Snr+_1}, we have
\[
\| \bX - \bW \|_\fro^2 \leq {\revR1color\sum_{i=1}^n \rho_1(\bar{\bx}_i)^2} \leq \| \bX \bone \|_2^2 - \| \bX \|_\fro^2.
\]

%-------
\bibliographystyle{ieeetr}
\bibliography{refs_JB_part2}

\ifconfver
\newpage
\begin{IEEEbiography}[{\includegraphics[width=1.2in,height=1.25in,clip,keepaspectratio]{./bios/Junbin Liu/ljb.png}}]{Junbin Liu} received the B.S. degree
from the South China University of Technology, Guangzhou, China, in 2017 and M.S. degree from the University of Chinese Academy of Sciences in 2020.
He is currently pursuing the Ph.D. degree
with the Department of Electronic Engineering, the Chinese University of Hong Kong, under the supervision of Professor Wing-Kin Ma.

His research interests encompass statistical signal processing methods, optimization theories, and their wide-ranging applications.
\end{IEEEbiography}

\begin{IEEEbiography}[{\includegraphics[width=1.2in,height=1.25in,clip,keepaspectratio]{./bios/Ya Liu/ly.jpg}}]{Ya Liu} received the B.S. degree from the University of Electronic Science and Technology of China (UESTC), Chengdu, China, in 2018.
She is currently pursuing a Ph.D. degree with the Department of Electronic Engineering at the Chinese University of Hong Kong (CUHK), under the supervision of Professor Wing-Kin Ma.

Her research focuses include matrix factorization, signal processing, and optimization, along with their diverse applications.
\end{IEEEbiography}

\begin{IEEEbiography}[{\includegraphics[width=1in,height=1.25in,clip,keepaspectratio]{./bios/Wing-Kin Ma/wingkinma.jpg}}]{Wing-Kin Ma}
 (Fellow, IEEE) is currently a Professor with the Department of Electronic Engineering, The
Chinese University of Hong Kong (CUHK), Hong Kong. His research interests include signal processing, optimization and communications, with recent focus on i) optimization and statistical aspects with structured matrix factorization, with application to remote sensing and data science; and ii) coarsely quantized MIMO transceiver designs.

Dr. Ma has rich experience in editorial service, such as an Associate Editor, and then later, a Senior Area Editor, and then from 2021 to 2023, the Editor-in-Chief of the
{\sc IEEE Transactions on Signal Processing},
%the Guest Editor or Lead Guest Editor of three special issues in {\sc IEEE Signal Processing Magazine}, 
and many others. He was a Tutorial Speaker in EUSIPCO 2011 and ICASSP 2014, and  was an IEEE Signal Processing Society (SPS) Distinguished Lecturer in 2018--2019. 	 
He was the recipient of the Research Excellence Award 2013--2014 by CUHK, the 2015 IEEE Signal Processing Magazine Best Paper Award, the 2016 IEEE Signal Processing
Letters Best Paper Award, and the 2018 IEEE SPS
Best Paper Award. He served as a Member of the Signal Processing
for Communications and Networking Technical Committee (SPCOM-TC) in 2015--2020,
a Member of 
Signal Processing Theory and Methods Technical Committee (SPTM-TC) in 2012--2017, 
the SPS Regional
Director-at-Large for Region 10 in 2020--2021, and a Technical Program Co-Chair of ICASSP 2023.
He co-founded and co-organized One World Signal Processing in 2020, a virtual seminar series for signal processing.
	
\end{IEEEbiography}

\begin{IEEEbiography}[{\includegraphics[width=1.2in,height=1.25in,clip,keepaspectratio]{./bios/Mingjie Shao/MingjieShao.jpg}}]{Mingjie Shao}
(S’16-M'20) received the B.S. degree
from the Xidian University, Xi'an, China, in 2015 and Ph.D. degree from the Chinese University of Hong Kong (CUHK) in 2020.
He was a Postdoctoral Fellow
with the Department of Electronic Engineering, CUHK from 2020 to 2023.
He is currently a  Research Professor (Qilu Young Scholar) in the School of Information Science and Engineering, Shandong University, Qingdao, China.
He was the recipient of the Hong Kong PhD Fellowship Scheme (HKPFS) from August 2015.
He was listed in the Student Best Paper Finalists in ICASSP 2017.

His research interests
focus on convex and non-convex
optimization, statistical signal processing and machine learning for wireless communication.

\end{IEEEbiography}

\begin{IEEEbiography}[{\includegraphics[width=1in,height=1.25in,clip,keepaspectratio]{bios/Anthony Man-Cho So/AnthonySo.jpg}}]{Anthony Man-Cho So} (Fellow, IEEE) received the B.S.E. degree in computer science from Princeton University, Princeton, NJ, USA, with minors in applied and computational mathematics, engineering and management systems, and German language and culture; the M.Sc. degree in computer science and the Ph.D. degree in computer science with a Ph.D. minor in mathematics from Stanford University, Stanford, CA, USA. He is currently Dean of the Graduate School, Deputy Master of Morningside College, and a Professor with the Department of Systems Engineering and Engineering Management at The Chinese University of Hong Kong (CUHK), Hong Kong SAR, China. His research interests include optimization theory and its applications in various areas of science and engineering, including computational geometry, machine learning, signal processing, and statistics. 
	
Dr. So is a Fellow of the Hong Kong Institution of Engineers. He is the recipient of a number of research and teaching awards, including the 2024 INFORMS Computing Society Prize, the SIAM Review SIGEST Award in 2024, the 2022 University Grants Committee Teaching Award, the 2018 IEEE Signal Processing Society Best Paper Award, the 2015 IEEE Signal Processing Society Signal Processing Magazine Best Paper Award, the 2014 IEEE Communications Society Asia-Pacific Outstanding Paper Award, the 2013 CUHK Vice-Chancellor’s Exemplary Teaching Award, and the 2010 INFORMS Optimization Society Optimization Prize for Young Researchers. He currently serves on the Editorial Boards of \emph{Journal of Global Optimization}, \emph{Mathematics of Operations Research}, \emph{Mathematical Programming}, \emph{Open Journal of Mathematical Optimization}, \emph{Optimization Methods and Software}, and \emph{SIAM Journal on Optimization}. He was also the Lead Guest Editor of the Special Issue on Non-Convex Optimization for Signal Processing and Machine Learning of the {\sc IEEE Signal Processing Magazine} and a Guest Editor of the Special Issue on Advanced Optimization Theory and Algorithms for Next Generation Wireless Communication Networks of the {\sc IEEE Journal on Selected Areas in Communications}.
\end{IEEEbiography}

\fi

\end{document}